%% file: main.tex
\documentclass[11pt,tightenlines,onecolumn,preprintnumbers,superscriptaddress,nofootinbib]{revtex4}

\usepackage[a4paper,left=2.5cm,right=2.5cm, top=1.5cm,bottom=1.5cm]{geometry}
\usepackage[english]{babel}
\usepackage[applemac]{inputenc}

\usepackage{amsfonts, amsmath, amssymb}
\usepackage{graphicx}
\usepackage{booktabs, multirow}
\usepackage{hyperxmp}
\usepackage{bbold}
\usepackage{rotating}
\usepackage{hyperref}
\hypersetup{colorlinks=true,citecolor=blue,linkcolor=cyan}

\newcommand{\fm}{\mathrm{fm}}
\newcommand{\MeV}{\mathrm{MeV}}

\newcommand{\psib}{\overline{\psi}}	

\newcommand{\ext}{\mathrm{ext}}	
\newcommand{\Tr}{{\rm Tr}}

\newcommand{\be}{\begin{equation}}
\newcommand{\ee}{\end{equation}}
\newcommand{\bea}{\begin{eqnarray}}
\newcommand{\eea}{\end{eqnarray}}
\newcommand{\bdm}{\begin{displaymath}}
\newcommand{\edm}{\end{displaymath}}

\newcommand{\mqav}{m_{\rm q}^{\rm av}}
\newcommand{\mql}{m_{{\rm q},l}}
\newcommand{\mqs}{m_{{\rm q},s}}
\newcommand{\cv}{c_{\rm V}}
\newcommand{\bv}{b_{\rm V}}
\newcommand{\bg}{b_{\rm g}}
\newcommand{\fv}{f_{\rm V}}
\newcommand{\bbar}{\overline{b}}
\newcommand{\bbarv}{\overline{b}_{\rm V}}
\newcommand{\zv}{Z_{\rm V}}
\newcommand{\za}{Z_{\rm A}}
\newcommand{\zp}{Z_{\rm P}}
\newcommand{\ca}{c_{\rm A}}
\newcommand{\ba}{b_{\rm A}}
\newcommand{\bbara}{\overline{b}_{\rm A}}
\newcommand{\mpcac}{m}
\newcommand{\Nc}{N_{\rm c}}
\newcommand{\Nf}{N_{\rm f}}
\newcommand{\<}{\langle}
\renewcommand{\>}{\rangle}
\newcommand{\hatp}{\hat{p}}
\newcommand{\circp}{p^{^{\!\!\!\circ}}}

\newcommand{\circk}{k^{^{\!\!\!\circ}}}
\newcommand{\circl}{\ell^{^{\!\!\!\circ}}}
\newcommand{\circvecq}{{\vec q}^{^{\!\!\!\circ}}}
\newcommand{\circveck}{{\vec k}^{^{\!\!\!\circ}}}
\newcommand{\circvecl}{{\vec\ell}^{^{\!\!\!\circ}}}

\renewcommand{\vec}[1]{\boldsymbol{#1}}

\makeatletter
\g@addto@macro\bfseries{\boldmath}
\g@addto@macro\normalfont{\unboldmath}
\makeatother

\begin{document}

\title{Non-perturbative renormalization and O$(a)$-improvement of the non-singlet vector current 
with $\Nf=2+1$ Wilson fermions and tree-level Symanzik improved gauge action}

\author{Antoine G\'erardin}
\affiliation{Institut f\"ur Kernphysik \& Cluster of Excellence PRISMA,  Johannes Gutenberg-Universit\"at Mainz, D-55099 Mainz, Germany}
\affiliation{John von Neumann Institute for Computing, DESY, Platanenallee 6, D-15738 Zeuthen, Germany}
\author{Tim Harris} 
\affiliation{Helmholtz Institut Mainz,  D-55099 Mainz, Germany}
\affiliation{
Dipartimento di Fisica  G. Occhialini,
Universita' di  Milano-Bicocca,
and,
INFN, Sezione di Milano-Bicocca,
Piazza della Scienza 3, I-20126 Milano, Italy}
\author{Harvey B.\ Meyer} 
\affiliation{Institut f\"ur Kernphysik \& Cluster of Excellence PRISMA,  Johannes Gutenberg-Universit\"at Mainz, D-55099 Mainz, Germany}
\affiliation{Helmholtz Institut Mainz,  D-55099 Mainz, Germany}

\date{\today}

\preprint{MITP/18-106}

\begin{abstract}
In calculating hadronic contributions to precision observables for tests of the Standard Model in lattice QCD, 
the electromagnetic current plays a central role. Using a Wilson action with O($a$) improvement
in QCD with $\Nf$ flavors, a counterterm must be added to the vector current in order for its on-shell matrix elements to be 
O($a$) improved. In addition, the local vector current, which has support on one lattice site, must be renormalized. 
At O($a$), the breaking of the SU($\Nf$) symmetry by the quark mass matrix 
leads to a mixing between the local currents of different quark flavors.
We present a non-perturbative calculation of all the required improvement and renormalization constants 
needed for the local and the conserved electromagnetic current 
in QCD with $\Nf=2+1$ O($a$)-improved Wilson fermions and tree-level Symanzik improved gauge action,
with the exception of one coefficient, which we show to be order $g_0^6$ in lattice perturbation theory.
The method is based on the vector and axial Ward identities imposed at finite lattice spacing and in the chiral limit.
We make use of lattice ensembles generated as part of the Coordinated Lattice Simulations (CLS) initiative.
\end{abstract}

\maketitle

\section{Introduction}

Precision tests of the Standard Model typically require reliable
theory input from first-principles calculations.  While the
electroweak sector can be treated perturbatively, the virtual
contributions of hadrons must be calculated from QCD
non-perturbatively. Ab initio Monte-Carlo simulations of lattice QCD
have provided a host of precision quantities for use in tests of the
Standard Model~\cite{Aoki:2016frl}.  Example of such hadronic quantities are the ratio of
decay constants $f_K/f_\pi$, the $\overline{\rm MS}$ quark masses, the
running strong coupling constant $\alpha_s(M_Z)$, and the anomalous
magnetic moment of the muon, $(g-2)_\mu$. For the latter,
a major effort by several lattice collaborations worldwide is ongoing
to calculate the hadronic vacuum polarization and the hadronic
light-by-light contributions~\cite{Meyer:2018til}. From the QCD point of view, these
contributions amount to two- and four-point correlation functions of
the electromagnetic current, to be integrated over with a weight
function containing the characteristic scale of the muon mass.

In continuum QCD, the electromagnetic current is conserved and does
not require renormalization. On the lattice, a finite renormalization
can appear, depending on the details of the action and of the chosen
discretization of the vector current. In particular, for Wilson
fermions, the single O($a$) on-shell improvement term to the action is
known.  Wilson fermions also have a `point-split' vector current,
whose support extends over two lattice sites in the direction of the
current, which is exactly conserved at finite lattice spacing. This
appealing property, however, does not guarantee that transverse
correlation functions of the current have smaller discretization
effects than those of the naive, ``local'' vector current with support
on a single lattice site, which in the limit of massless quarks
receives a finite renormalization factor $\zv(g_0^2)$.  Indeed, the
improvement of the vector current -- local or point-split -- requires
adding the divergence of the tensor current with a coefficient denoted
$\cv$, which counteracts the breaking of chiral symmetry by the Wilson action 
and suffices to remove all O($a$) cutoff effects in on-shell
correlation functions. This coefficient, whose value depends on the discretization of the current, has a finite value at
tree level of perturbation theory in the case of the point-split current, but
vanishes for the local current.

On the other hand, for the local vector current, a mass dependence of
the renormalization factor arises if O($a$) discretization errors are
to be removed.  This mass dependence is relevant in precision
calculations, given the pattern of the physical up, down and strange
quark masses. Concretely, given the electric charge matrix of the
lightest three quark flavors, ${\cal Q}={\rm diag}(2/3,-1/3,-1/3)$,
the electromagnetic current can be written as the linear combination
\be
V_\mu^{\rm e.m.} = V_\mu^3 + \frac{1}{\sqrt{3}} V_\mu^8\,,
\ee
where $V_\mu^a = \bar\psi \gamma_\mu \frac{\lambda^a}{2}\psi$ is the octet of vector currents, 
with $\lambda^a$ the Gell-Mann matrices. In isospin-symmetric QCD, the bare quark mass matrix can be decomposed as 
\be\label{eq:Mq}
M_{\rm q} = \mqav + \frac{1}{\sqrt{3}}(m_{{\rm q},l}-m_{{\rm q},s}) \lambda^8 \,.
\ee
See Eq.~(\ref{eq:m_q}) and below for our notation. The renormalization pattern of the local discretization of the 
two neutral octet combinations then reads~\cite{Bhattacharya:2005rb}, at O($a$),
\bea\label{eq:Vmu3}
V_{\mu,R}^{3} &=& \zv\; (1+ 3\bbarv\; a\mqav + \bv\;a\mql) \;V_\mu^{3,I}\,,
\\
\label{eq:Vmu8}
V_{\mu,R}^{8} &=& \zv\; \bigg[ \Big(1+ 3 \bbarv\; a\mqav + \frac{\bv}{3}\; a(\mql+2\mqs)\Big) \,V_\mu^{8,I} 
\nonumber    \\ && \qquad ~       +\, ({\textstyle\frac{1}{3}}\bv+\fv)\; \frac{2}{\sqrt{3}}a(\mql-\mqs) \;V_\mu^{0,I}\bigg]\,,
\eea
with $V_\mu^{0} = \frac{1}{2}\bar\psi \gamma_\mu \psi$ the flavor-singlet current.
Here $V_\mu^{3,I}$ and $V_\mu^{8,I}$ are understood to already contain the improvement term proportional to $\cv$.
All coefficients appearing in the two equations above are functions of the coupling $\widetilde{g}_0$. 
In this article, we present a non-perturbative determination of the
renormalization factors $\zv$, $\bv$ and $\bbarv$ as well as of
$\cv$, while the coefficient $\fv$ will remain undetermined. 
As explained below, there are, however, good reasons to expect $\fv$ to be
numerically very small~\cite{Bhattacharya:2005rb}.
The improvement coefficient $\cv$ is determined by imposing continuum chiral Ward identities, as proposed in quenched QCD in Ref.~\cite{Guagnelli:1997db}.
We follow the presentation of Ref.~\cite{Bhattacharya:2005rb} for the full renormalization and improvement in large volumes with $N_\mathrm{f}=2+1$ Wilson fermions.
The mass-dependent renormalization with $N_\mathrm{f}=2$ Wilson fermions has been computed in Ref.~\cite{Bakeyev:2003ff}.
Note that the method of Ref.~\cite{Harris:2015vfa} allows only one to compute a linear combination 
of the improvement coefficients for the conserved and local currents, and it is insufficient to provide a full improvement condition for either discretization.

Our main motivation for the present calculation is to determine the
two-point function of the electromagnetic current with only O($a^2$)
discretization effects. This will in particular allow for a shorter
continuum extrapolation of the leading hadronic contribution to the
anomalous magnetic moment of the muon, and therefore a more
cost-effective set of lattice QCD simulations. Given that phenomenologically, 
the $\pi^+\pi^-$ channel, which is described by the timelike electromagnetic  form factor of the pion,
accounts for more than two-thirds of the total hadronic contributions, it is very 
natural to impose the renormalization condition on the local vector current 
that the electric charge of the pion be unity at every lattice spacing. This 
is the main renormalization condition we will adopt to determine $\zv$, $\bv$ and $\bbarv$.

We begin by giving an overview of the required theory background,
which allows us to define our notation. We present the setup for our
numerical calculation in Sec.~\ref{sec:num_setup} and the results
in Sec.~\ref{sec:results}. We finish with some concluding remarks
in Sec.~\ref{sec:concl}.  Appendix \ref{app:C} presents a
determination of the improvement coefficient $\ca$ of the axial
current, and Appendix \ref{app:B} contains some results on the
employed correlation functions in lattice perturbation theory.

\section{Renormalization and improvement: theory background}

\subsection{Definitions and notations} 

We use Euclidean Dirac matrices, $\{\gamma_\mu,\gamma_\nu\}=2\delta_{\mu\nu}$.
We consider initially the general case of $\Nf$ flavors of quarks.
Flavor indices will be denoted by latin letters $i,j,\dots$
Let 
\begin{equation}
A_{\mu}^{(ij)}(x) =   \psib_i(x) \gamma_{\mu} \gamma_5 \psi_j(x) \,, \quad  P^{(ij)}(x) = \psib_i(x) \gamma_5 \psi_j(x) \,
\end{equation}
be the bare axial current and pseudoscalar density.
The on-shell improved operators are given by 
\begin{equation}
(A_I^{(ij)})_{\mu}(x) = A_{\mu}^{(ij)}(x) + a\ca(g_0^2) \, \partial_{\mu} P^{(ij)}(x) \,, \quad P_I^{(ij)}(x) = P^{(ij)}(x) 
\qquad (i\neq j)\,,
\label{eq:imp_op}
\end{equation}
where $\ca$ is an improvement coefficient. The average bare PCAC quark mass $\mpcac_{ij}$ of quark flavors $i$ and $j$ is defined through the relation
\begin{equation}
\langle \partial_{\mu} (A^{(ij)}_{I})_{\mu}(x) \, P^{(ji)}(y) \rangle = 2 \mpcac_{ij}\, \langle P_I^{(ij)}(x) \, P^{(ji)}(y) \rangle 
+ {\rm O}(a^2) \qquad (i\neq j,x\neq y)\,.
\label{eq:def_PCAC}
\end{equation}
We also defined the subtracted bare quark mass of flavor $i$,
\begin{equation}
m_{{\rm q},i} =  m_{0,i} - m_{\rm cr}.
\label{eq:m_q}
\end{equation}
Often, the hopping parameter  $\kappa_i \equiv (8+2am_{0,i})^{-1}$ is used to parametrize the bare quark mass $m_{0,i}$.
The value $\kappa_{\rm cr}\equiv (8+2am_{\rm cr})^{-1}$ of the hopping parameter is the value for which the PCAC mass, defined through Eq.~(\ref{eq:def_PCAC}), vanishes in the ${\rm SU}(\Nf)$-symmetric theory. 
The bare quark mass matrix is defined as $M_0={\rm diag}(m_{0,1},\dots,m_{0,\Nf})$, and similarly for the subtracted bare quark mass matrix, $M_{\rm q}={\rm diag}(m_{{\rm q},1},\dots,m_{{\rm q},\Nf})$. 
Finally, we also introduce the average quark masses
\begin{equation}
m_{{\rm q},ij} = {\textstyle\frac{1}{2}} (m_{{\rm q},i} + m_{{\rm q},j} )\,, \quad \mqav = \frac{1}{\Nf} \sum_{i=1}^{\Nf} m_{\rm q,i} \,.
\end{equation}

Here we will be concerned with the improvement and renormalization of the vector current $V_\mu^{(ij)}$ on the lattice.
Two discretizations are in common use,  the local $(l)$ and the point-split $(c)$ lattice vector currents,
\begin{subequations}
\begin{align}
V_{\mu}^{l,(ij)}(x) &= \psib_{i}(x) \gamma_{\mu} \psi_j(x) \,,\\
V_{\mu}^{c,(ij)}(x) &= \frac{1}{2} \left( \psib_i(x+a\hat{\mu})(1+\gamma_{\mu}) U^{\dag}_{\mu}(x) \psi_j(x) - \psib_i(x) (1-\gamma_{\mu} ) U_{\mu}(x) \psi_j(x+a\hat{\mu}) \right) \,.
\label{eq:consvec1}
\end{align}  
\end{subequations}
Instead of the point-split vector current, we actually consider the symmetrized version $(cs)$
\begin{equation}
V_{\mu}^{cs,(ij)}(x) = \frac{1}{2} \left( V_{\mu}^{c,(ij)}(x) + V_{\mu}^{c,(ij)}(x-a\hat{\mu})  \right) \,,
\end{equation}
which has the same properties under spacetime reflections as the local vector current\footnote{The authors thank Stefan Sint for pointing out this fact}~\cite{Frezzotti:2001ea}. This ensures that the same counterterms are present to remove O($a$) artifacts
\begin{equation}
\label{eq:imp2}
(V_I^{(ij)})_{\mu}(x) = V_{\mu}^{(ij)}(x) + a\cv(g_0^2) \, \widetilde\nabla_{\nu} \Sigma_{\mu\nu}^{(ij)}(x) \,,
\end{equation}
with the local tensor current defined as 
\be
\Sigma_{\mu\nu}^{(ij)} = -\frac{1}{2} \psib_i  [\gamma_{\mu}, \gamma_{\nu}] \psi_j \,,
\ee
and where we use the symmetric lattice derivative,
\begin{equation}
\widetilde\nabla_{\nu} \phi(x) = \frac{ \phi(x+a\hat{\nu}) - \phi(x-a\hat{\nu})  }{2a} \,.
\label{eq:defT}
\end{equation}
Generically, the renormalization pattern of the quark bilinears, including O($a$) mass-dependent effects,
has been derived in Ref.~\cite{Bhattacharya:2005rb}. For the vector current, and writing $V_{\mu}$ as a flavor matrix, it reads
\bea
\label{eq:imp}
\mathrm{tr}( \lambda V_{\mu} )_R &=&
 \zv(\tilde g_0^2) \Big[ \left(1+ \Nf\,\bbarv(g_0^2) \, a\mqav \right) \mathrm{tr}( \lambda V_{\mu}^{I} )  
  + \frac{1}{2} \bv(g_0^2) \, \mathrm{tr}( \{ \lambda, aM_{\rm q} \} V_{\mu}^{I} ) 
\nonumber\\ && \qquad \qquad +\;  \fv(g_0^2) \, \mathrm{tr}( \lambda\, aM_{\rm q} ) \, \mathrm{tr}(V_{\mu}^{I}) \Big] \,, 
\eea
where 
\be
\tilde g_0^2 \equiv g_0^2 (1+\bg\,a\mqav)
\ee
is the modified bare coupling, which is in one-to-one correspondence with the lattice spacing, irrespective of the quark masses~\cite{Luscher:1996sc}. The symbol `$\mathrm{tr}$' refers to the trace over flavor indices and $\lambda$ is any element of the SU($\Nf$) Lie algebra. The improvement coefficients $\cv$, $\bv$, $\bbarv$ and $\fv$ are functions of the bare coupling only; $\zv$ has no
anomalous dimension and does not depend on the renormalization scale. 

Given that the coefficient $\bg$ is so far only known perturbatively, it is worth noting  the following. 
If one Taylor expands the function $\zv$ and only keeps terms up to O($a$), the expression~(\ref{eq:imp}) is equivalent to replacing the argument of $\zv$ by $g_0^2$ and then substituting $\bbarv$ by
\be\label{eq:bbareff_def}
\bbarv^{\,\rm eff}(g_0^2) \equiv \bbarv(g_0^2)+{\frac{1}{\Nf}} \bg(g_0^2)\, \frac{g_0^2}{\zv}\frac{d\zv}{dg_0^2} .
\ee
Therefore, the renormalization conditions we use for the vector
current are only able to determine the combination
$\bbarv^{\,\rm eff}$. In a second step, using the perturbative
estimate of $\bg$, we obtain a value for $\bbarv$.
In the future, when a non-perturbative determination of $\bg$ becomes available,
the value of $\bbarv$ can be updated.

In Sec.~\ref{sec:VWI}, we describe the strategy used to determine
the renormalization constant $\zv$ and the improvement coefficients
$\bv$, $\bbarv^{\,\rm eff}$ and $\cv$.

\subsection{Vector Ward identities and determination of $\zv$, $\bv$ and $\bbarv$ \label{sec:VWI}}

We define an infinitesimal local vector transformation by 
\begin{subequations}
\begin{align}
\delta \psi(y) &= \lambda\,\alpha(y) \,\psi(y) \,, \\
\delta\overline{\psi}(y) &= - \overline{\psi}(y) \alpha(y)\, \lambda \,,
\label{eq:VT}
\end{align}
\end{subequations}
where the  matrix $\lambda$ acts on flavor space. Using the path integral definition of an expectation value and noticing 
that the previous transformation is a change of integration variables with unit Jacobian, one obtains the following identity,
\begin{equation}
\Big\langle \frac{ \delta \mathcal{O} }{ \delta \alpha(y) } \Big\rangle 
= \Big\langle \mathcal{O} \frac{ \delta S}{ \delta \alpha(y)} \Big\rangle \,,
\end{equation}
where $S$ is the Euclidean action and $\mathcal{O}$ is any operator. 
In fact, the equality holds on every single gauge-field configuration because only the fermionic part of the action is affected.
For Wilson-Clover fermions, it leads to the well-known vector Ward identity~\cite{Bochicchio:1985xa}
\begin{equation}
\Big \langle \frac{ \delta \mathcal{O} }{ \delta \alpha(y) } \Big\rangle 
= a^4 {\nabla}_{\mu}^* \Big\langle {\rm tr}\{\lambda^\intercal \,V_{\mu}^{c}(y)\} \,\mathcal{O} \Big\rangle 
+ a^4 \Big\langle \psib(y) \left[ M_0,\, \lambda \right] \psi(y) \, \mathcal{O} \Big\rangle   \,,
\label{eq:WI}
\end{equation}
where ${\nabla}_{\mu}^* \phi(y) = (\phi(y)-\phi(y-a\hat{\mu}))/a$ is the backward lattice derivative in the $\mu$-direction,
$\lambda^\intercal$ denotes the matrix transpose of $\lambda$ and 
$ V_\mu^{c,(ij)}(y)$ corresponds to the point-split vector current defined in Eq.~(\ref{eq:consvec1}).
Using an operator $\mathcal{O}$ with support which does not contain the site $y$ 
and for $\left[ M, {\lambda}  \right]=0$, one simply recovers the conservation equation for the point-split vector current.

Working in components, we now consider  the vector transformation 
\begin{align}
\delta \psi_i(y) = + \alpha(y) \psi_i(y) \,, \quad \delta \psib_i(y) = - \alpha(y) \psib_i(y) 
\end{align}
for one specific flavor $i$.
Then, using $\mathcal{O}(x,z) = P^{(ji)}(x)P^{(ij)}(z)$ as a probe operator with $i\neq j$, one finds
\begin{equation}
\frac{ \delta \mathcal{O}(x,z) }{\delta \alpha(y)} =  \mathcal{O}(x,z) \delta(y-x) -  \mathcal{O}(x,z) \delta(y-z) \,,
\end{equation}
such that Eq.~(\ref{eq:WI}) reads
\begin{equation}
\Big\langle P^{(ji)}(x)P^{(ij)}(z) \Big\rangle \left( \delta(y-z) - \delta(y-x) \right) = a^4 {\nabla}_{y,\mu}^*
\Big\langle  P^{(ji)}(x) V_{\mu}^{c,(ii)}(y) P^{(ij)}(z)  \Big\rangle  \,.
\label{eq:wi_final}
\end{equation}
Summing over the spatial vector $\vec y$ in Eq.~(\ref{eq:wi_final}), 
the spatial derivative does not contribute due to the use of periodic boundary conditions 
and only the time derivative remains. Therefore, the  three-point correlation function
 $\langle a^3\sum_{\vec{y}} P^{(ji)}(x) V_{0}^{c,(ii)}(y) P^{(ij)}(z) \rangle$, viewed as a function of $y_0$,
is a piecewise constant function with discrete steps of $+1$ at $y_0=z_0$ and $-1$ at $y^0=x^0$. 
In particular, for $x_0 > y_0 > z_0$, the ratio $R$ defined by
\begin{equation}
R(x_0-z_0, y_0-z_0) = \frac{ \langle a^6\sum_{\vec{x},\vec{y}} P^{(ji)}(x) V_{0}^{c,(ii)}(y) P^{(ij)}(z) \rangle }
{ \langle a^3\sum_{\vec{x}} P^{(ji)}(x) P^{(ij)}(z) \rangle} \,,
\label{eq:condition}
\end{equation}
is unity such that the point-split vector current does not need any
renormalization factor : $\zv^c=1$ and $\bbarv^c=\bv^c=\fv^c=0$. On
the other hand, the local vector current is not conserved on the
lattice and needs to be renormalized.

In $\Nf=2+1$ QCD with a quark mass matrix given by (\ref{eq:Mq}), by imposing either of the ratios
\begin{subequations}
\begin{align}
R_\pi(x_0-z_0, y_0-z_0) &= \frac{ \langle a^6\sum_{\vec{x},\vec{y}} 
P^{(21)}(x) \,\frac{1}{2}(V_{0,\mathrm{R}}^{l,(11)}(y)-V_{0,\mathrm{R}}^{l,(22)}(y)) P^{(12)}(z) \rangle }
{ \langle a^3\sum_{\vec{x}} P^{(21)}(x) P^{(12)}(z) \rangle} \,,
\\
R_K(x_0-z_0, y_0-z_0) &= \frac{ \langle a^6\sum_{\vec{x},\vec{y}} 
P^{(31)}(x) \,(V_{0,\mathrm{R}}^{l,(11)}(y)-V_{0,\mathrm{R}}^{l,(22)}(y)) P^{(13)}(z) \rangle }
{ \langle a^3\sum_{\vec{x}} P^{(31)}(x) P^{(13)}(z) \rangle} 
\end{align}
\end{subequations}
to equal unity on the lattice at finite quark masses, one
can determine the renormalization factor of the local isovector current 
$V_\mu^3 = \frac{1}{2}(V_\mu^{(11)}-V_\mu^{(22)})$,
including the O($a$) mass-dependent terms, as given explicitly in Eq.~(\ref{eq:Vmu3}).
We note that this renormalization condition does not require the
knowledge of $\cv$, and that the two choices  for the ``spectator
quark''  correspond to two different renormalization prescriptions.
Using  ensembles with different values of $m_{\rm q,1}=m_{{\rm q},2}$ and $m_{\rm q,3}=m_{{\rm q},s}$,
each parameter can be determined independently. 
We remark that $\zv$, $\bv$ and $\bbarv^{\,\rm eff}$ could also be determined in the same way from
the matrix element of 
\be
\widetilde R_K(x_0-z_0, y_0-z_0) = 
\frac{ -\langle a^6\sum_{\vec{x},\vec{y}} 
P^{(31)}(x) \,\frac{1}{3}(V_{0,\mathrm{R}}^{l,(11)}(y)+V_{0,\mathrm{R}}^{l,(22)}(y)-2V_{0,\mathrm{R}}^{l,(33)}(y)) P^{(13)}(z) \rangle }
{ \langle a^3\sum_{\vec{x}} P^{(31)}(x) P^{(13)}(z) \rangle},
\ee
since the flavor-singlet charge operator does not contribute on a kaon state.
On the other hand, to obtain sensitivity to the coefficient $\fv$,
an external state with nonvanishing baryon number is required, for instance
one may require the ratio
\be
 R_{\Delta^{++}}(x_0-z_0, y_0-z_0) = 
 \frac{ \langle a^6\sum_{\vec{x},\vec{y}} 
  \Delta^{(111)}(x) \,(V_{0,\mathrm{R}}^{l,(22)}(y)-V_{0,\mathrm{R}}^{l,(33)}(y)) \bar\Delta^{(111)}(z) \rangle }
 { \langle a^3\sum_{\vec{x}} \Delta^{(111)}(x) \bar\Delta^{(111)}(z) \rangle}
\ee
to vanish.
Without the vector current improvement term proportional to $\fv$, $R_{\Delta^{++}}$ would receive a contribution 
of order $a$ from disconnected diagrams; the role of the coefficient $\fv$, which multiplies the flavor-singlet vector current, under which the $\Delta^{++}$ baryon is charged,  is to cancel this contribution. Therefore, the magnitude of $\fv$ is determined by the size of disconnected diagrams with the insertion of a single vector current. In perturbation theory, $\fv$ is therefore of order $g_0^6$, because at least three gluons must be emitted from the quark loop\footnote{One-gluon exchange does not contribute because the color factor vanishes.
To see that the two-gluon exchange also vanishes, one may use the $\gamma_5$-hermiticity of the quark propagator, $\gamma_5 S(x,y)\gamma_5 = S(y,x)^\dagger$, the fact that the free quark propagator $S(x,y)$ is Hermitian for fixed $(x,y)$ and $\gamma_5 \gamma_\mu \gamma_5 = -\gamma_\mu$, to show that the two orientations with which the quark propagators contribute to the quark loop come with opposite signs and cancel each other.} 

\subsection{Axial Ward identities and determination of $\cv$}
\label{sec:cv}

\begin{figure}[t]
	\includegraphics*[width=0.9\linewidth]{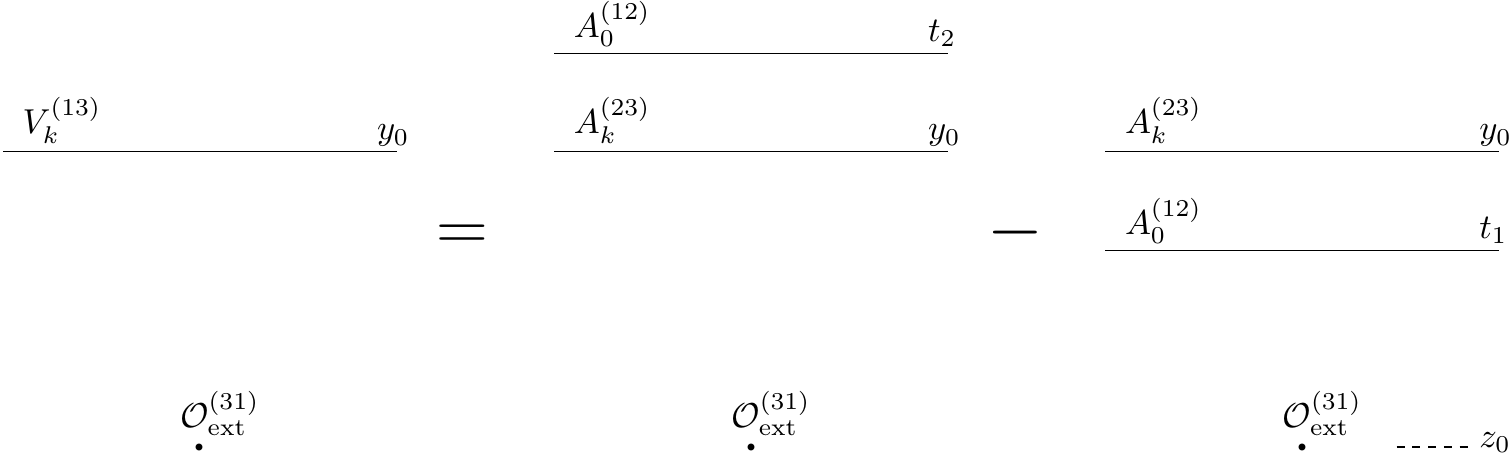}
\caption{The chiral Ward identity in the continuum and in the limit $m_{12}=0$.
Continuous horizontal lines indicate that the operator is projected onto vanishing spatial momentum.
\label{fig:WIpic}}
\end{figure}

Once the renormalization factor $\zv$ and improvement coefficients $\bv$ and $\bbarv^{\,\rm eff}$ are known, the improvement coefficient $\cv$ can be determined by enforcing an axial Ward identity. In the continuum theory, the latter can be derived from  the specific transformation
\be
\delta\psi_1(x) = -\alpha(x)\gamma_5 \psi_2(x)\,, \qquad \delta\overline\psi_2(x) = -\overline\psi_1(x)\alpha(x)\gamma_5 \,.
\ee
As the operator to be chirally rotated, we choose  ${\cal O}(y) = A_\mu^{(23)}(y)$, and we have 
\be
\delta A_\mu^{(23)}(y) = \alpha(y) V_\mu^{(13)}(y)\,.
\ee
Choosing $\alpha(x)$ to be unity inside the slab $t_1 < x_0 < t_2$ and zero outside, the variation of the action (per unit $\alpha$) is given by 
\begin{equation}
\delta S^{(12)}
 =  -\int_{t_1}^{t_2} \mathrm{d} x^0 \, \int \mathrm{d}^3x \left( 2 \mpcac_{R,12} P_R^{(12)}(x) - \partial_{\mu} A_{R,\mu}^{(12)}(x) \right) \,,
\label{eq:dS}
\end{equation}
with $t_1 < y_0 < t_2$. We perform the integral over the divergence of the axial current explicitly in the continuum using Gauss's theorem. With the additional constraint $z_0 \notin [t_1,t_2]$, we then obtain the following Ward identity
\begin{equation}
\int \mathrm{d}^3y \, \Big\langle \delta S^{(12)} A_{R,k}^{(23)}(y_0,\vec{y}) \, \mathcal{O}_{\rm ext}^{(31)}(z_0,\vec 0) \Big\rangle 
= \int \mathrm{d}^3y \, \Big\langle V_{R,k}^{(13)}(y_0,\vec{y}) \, \mathcal{O}_{\rm ext}^{(31)}(z_0,\vec 0) \Big\rangle \,,
\label{eq:cor_cV}
\end{equation}
valid in the continuum~\cite{Guagnelli:1997db}, and impose it to hold on the lattice, at fixed quark mass and bare lattice coupling, up to higher order corrections O($a^2$). The Ward identity in the chiral limit is illustrated in Fig.~\ref{fig:WIpic}. For each discretization of the vector current ($\alpha=l,cs$), we define our estimator
\begin{equation}
\hat{c}^{\alpha}_{\mathrm{V}}(m_q,g_0^2) =  \frac{ \langle \sum_{ \vec{y}} 
\delta S^{(12)} A_{R,k}^{(23)}(y_0,\vec{y}) \, \mathcal{O}_{\rm ext}^{(31)}(z_0,\vec 0)  -  \hat{Z}_V^{(13)} \, \sum_{ \vec{y}}  V_{k}^{\alpha,(13)}(y_0,\vec{y}) \, \mathcal{O}_{\rm ext}^{(31)}(z_0,\vec 0)\rangle }
{ \hat{Z}_V^{(13)} \langle  \sum_{ \vec{y}} a \, \partial_{\nu} \Sigma_{k\nu}^{(13)}(y_0,\vec{y}) \, \mathcal{O}_{\rm ext}^{(31)}(z_0,\vec 0)\rangle } 
\label{eq:cV_lat}
\end{equation}
with 
\begin{equation}
\hat{Z}_V^{(13)}(g_0^2,\mqav,m_{\rm q,13}) = \zv(g_0^2) \left(1 + 3\bbarv^{\,\rm eff}(g_0^2) a \mqav + \bv(g_0^2) am_{\rm q,13} \right)
\label{eq:zhat13}
\end{equation} 
for the local vector current and $\hat{Z}_V^{(13)}(g_0^2,\mqav,m_{\rm q,13}) = 1$ for the conserved vector current. In Eq.~(\ref{eq:cV_lat}) we use the symmetric lattice derivative, as in Eq.~(\ref{eq:defT}). Other choices would differ by an O($a$) ambiguity in the definition of $\cv$. The renormalized axial and pseudoscalar operators for $i \neq j$ are, respectively, given by
\begin{subequations}
\begin{align}
P_R^{(ij)}(x) &= \zp \, ( 1 + 3 \bbar_{\rm P} \,a \mqav + b_{\rm P} \,am_{{\rm q},ij}) \ P_I^{(ij)}(x) \,, \\
A_{k,R}^{(ij)}(x) &= \za \, ( 1 + 3 \bbara \,a \mqav + \ba \,am_{{\rm q},ij}) \, A_{k,I}^{(ij)}(x) \,, 
\end{align}
\end{subequations}
in terms of the improved operators defined in Eq.~(\ref{eq:imp_op}). The renormalized quark mass is defined through the relation
\begin{equation}
\mpcac_{ij} =   \mpcac_{R,ij} \, \frac{\zp ( 1 + 3 \bbar_P \,a \mqav + b_{\rm P}\, a m_{{\rm q},ij} )}
{ \za (1 + 3 \bbara \,a \mqav  + \ba\, a m_{{\rm q},ij}) } \qquad (i\neq j)\,,
\end{equation}
such that the combination $ m_{R,12}P_R^{(12)}$ is insensitive to $\zp$, $b_{\rm P}$ and $\bbar_{\rm P}$.
The renormalization factor $\za$ and the improvement parameters $\ba$ and $\bbara$ have been determined 
non-perturbatively in Refs.~\cite{DallaBrida:2018tpn,Bulava:2016ktf,Korcyl:2016ugy}.

In Eq.~(\ref{eq:cV_lat}), the operator $\mathcal{O}_{\rm ext}$ can be either the vector $\mathcal{O}_{\rm ext}=V_{k}^{(31)}(z_0,\vec 0)$ 
or the tensor operator $\mathcal{O}_{\rm ext}=\Sigma_{k0}^{(31)}(z_0,\vec 0)$ and does not need to be O($a$)-improved. In perturbation theory, the choice of the tensor operator is superior, since in the continuum, the right-hand side of the Ward identity
vanishes in the chiral limit; on the lattice, the improvement term then involves the two-point function of the tensor current and its task is 
to cancel the vector-tensor correlation, which is O($a$) and originates from chiral symmetry breaking at the cutoff scale.
As we will see in the next section, in the non-perturbative QCD vacuum, both choices are equally well suited for separations
between the operators of order of 0.5\,fm, because the vector-tensor correlation is then nonvanishing even in the continuum.

There is one subtlety here. In Eq.~(\ref{eq:dS}), we sum
over all time slices in the range $[t_1,t_2]$ which implies the
presence of a contact term for $x_0 = y_0$. Therefore, on-shell
O($a$)-improvement is not sufficient to remove all
O($a$) contributions and the limit $\mpcac_{12} \to 0$ must
be taken to remove this contact term. This is done by computing the
effective $\hat{c}_V$ for different light quark masses using Eq.~(\ref{eq:cV_lat}) and then
extrapolating to the chiral limit.

Finally, in Appendix A we briefly describe a way to determine the improvement coefficient $\ca$ using 
an axial Ward identity. Our non-perturbative determination of $\ca$, which we can compare 
to the literature~\cite{Bulava:2015bxa}, serves as a cross-check of our numerical setup.

\subsection{Known perturbative results \label{sec:pert}}

The known perturbative results in QCD with $\Nc$ colors and $\Nf$ flavors of quarks  are the following. 
The result $\bg = 0.012000(2)\, \Nf\, g_0^2 + {\rm O}(g_0^4)$, independently of the pure gauge action, was obtained in~\cite{Sint:1995ch}.
For degenerate quarks, only the combination $\bv + \Nf\bbarv $ appears, and the perturbative series for $\bbarv$ starts at order $g_0^4$.
For the tree-level improved L\"uscher-Weisz action, the results are 
($C_{\rm F}= \frac{\Nc^2-1}{2\Nc}$)~\cite{Aoki:1998qd,Taniguchi:1998pf}
\begin{subequations}
\begin{eqnarray}
\zv &=& 1 -0.075427\times C_{\rm F}\, g_0^2 + {\rm O}(g_0^4)\,,
\\
\label{eq:bvpert}
\bv &=&  0.0886(2) \times  C_{\rm F}\, g_0^2 + {\rm O}(g_0^4)\,,
\\
\cv^l &=&   -0.01030( 4 )\times C_{\rm F} \, g_0^2 + {\rm O}(g_0^4) \,.
\end{eqnarray}
\end{subequations}
The tree-level value of $\cv^{cs}$ is $\frac{1}{2}$.


\input{table}

\section{Numerical setup \label{sec:num_setup}}

\begin{table}[b]
\caption{Values of $z_0$, $t_1,t_2$ and $y_0$ for the calculation of the three-point correlation function as defined in Eq.~(\ref{eq:cor_cV}). In the last column, we give the two values of $y_0$ used to interpolate to a line of constant physics as explained in the text. Note that ensembles at $\beta = 3.46$ were generated using periodic boundary condition in the time direction whereas other ensembles were generated using open boundary conditions. }
\vskip 0.1in
 \begin{tabular}{c@{\hskip 02em}c@{\hskip 02em}c@{\hskip 02em}c@{\hskip 02em}c}
	\hline
	$\beta$	&	$T/(2a)$	&	$z_0/a$	&	$[t_1,t_2]/a$	&	$y_0/a$	 \\
	\hline
	3.40		&	48		&	41	&	[46,54]		&	$49-50$	\\
	3.46		&	32		&	0	&	[6,15]		&	$10-11$	\\  
	 		&	48		&	0	&	[6,15]		&	$10-11$	\\  
	3.55		&	48		&	41	&	[48,59]		&	$52-53$	\\  
	 		&	64		&	57	&	[64,75]		&  	$68-69$	\\  
	3.70		&	64		&	53	&	[62,76]		&	$68-69$ 	\\  
	\hline
 \end{tabular} 
\label{tab:cV_setup}
\end{table}

We use the $N_f=2+1$ CLS (Coordinated Lattice Simulations) lattice ensembles~\cite{Bruno:2014jqa} whose main parameters are given in Table.~\ref{tab:sim}. 
They have been produced using the openQCD code\footnote{http://luscher.web.cern.ch/luscher/openQCD/} of Ref.~\cite{Luscher:2012av} using the Wilson-Clover action for fermions and the tree-level Symanzik improved gauge action. The parameter $c_{\rm SW}$ has been determined non-perturbatively in Ref.~\cite{Bulava:2013cta}. We consider four values of the bare coupling $\beta = 3.40, 3.46, 3.55$ and $3.70$ which correspond to lattice spacings in the range $0.050-0.085~\fm$~\cite{Bruno:2016plf}. Ensembles using (anti) periodic boundary conditions (PBC) and open-boundary conditions (OBC) in the time direction have been generated on three different chiral trajectories. Two trajectories with constant $\mqav$ and $\mqs=\mqs^{\rm phys}$ can be used to extrapolate results to the physical limit with physical masses of the light and strange quarks. A third trajectory uses degenerate light and strange quarks with $\mql = \mqs$.
Concerning $\cv$, it is enough to consider ensembles on a single chiral trajectory (e.g. $\mqav =$~const). However, to determine the
two improvement coefficients $\bv$ and $\bbarv^{\,\rm eff}$, we have to consider at least two different chiral trajectories. \\

For the calculation of the renormalization factor $\zv$, we need to compute the following three-point correlation function, projected on vanishing momentum
\begin{equation}
C_{PVP}(x_0,y_0;z_0) = a^6\sum_{\vec{x}, \vec{y}} 
\langle P^{(ij)}(x_0, \vec{x}) V^{(jj)}_0(y_0, \vec{y}) P^{(ij)\dag}(z_0, \vec{0}) \rangle \,,
\label{eq:3pt}
\end{equation}
and two-point correlation functions
\begin{subequations}
\begin{eqnarray}
C_{PP}(x_0,z_0) &=& a^3\sum_{\vec{x}} \langle P^{(ij)}(x_0, \vec{x}) P^{(ij) \dag}(z_0, \vec{0}) \rangle \,, 
\\
C_{AP}(x_0,z_0) &=& a^3\sum_{\vec{x}} \langle A^{(ij)}_0(x_0, \vec{x}) P^{(ij) \dag}(z_0, \vec{0}) \rangle \,.
\end{eqnarray}
\end{subequations}
Correlation functions are calculated using U(1) stochastic sources with time dilution~\cite{Foley:2005ac}. 
On each gauge configuration, we generate an ensemble of $N_s$ stochastic sources with support on a single time slice as well as satisfying
\begin{equation}
\lim_{N_s\rightarrow \infty} \frac{1}{N_s} \sum_{s=1}^{N_s} \eta_{\alpha}^a(x)_s \left[ \eta_{\beta}^b(y)_s \right]^{*} = a^{-3}\delta_{\alpha\beta} \delta^{ab}\delta_{x,y} \,,
\label{eq:stoch_sources}
\end{equation}
where each component is normalized to one, $\eta_{\alpha}^a(x)_{[r]}^{*}\ \eta_{\alpha}^a(x)_{[r]} = 1$ (no summation). This can be implemented by using U(1) noise for each color and spinor index on site $x$ of the lattice. For ensembles with open boundary conditions in the time direction, time translation is lost. In this case, the source is placed at $z_0=T/4$ away from the boundary ($t=0$) and the two-point correlation functions are obtained for all values of $x_0 \in [0, T/2]$, keeping the sink time away from the second boundary ($t=T$). For the three-point correlation function, the sink time is placed at $x_0=3T/4$ and is computed for all $y_0$.

For each stochastic source $s$ with support in time slice $z_0$, we solve the Dirac equation and denote the solution vector 
$\Phi_i^{s}(x;z_0)= a^3\sum_{\vec z} S(x,z) \eta_i^{s}(z)$. 
Correlation functions are given by
\begin{subequations}
\begin{align}
C^{(ij)}_{PP}(x_0, z_0) &= \frac{a^6}{V} \sum_{\vec{x}, \vec{z}} \Tr \left[ S_i(x,z)^{\dag} S_j(x,z)  \right] 
=  \frac{a^3}{N_s V} \sum_{s,\vec{x}}  \Phi_i^s(x;z_0)^{\dag} \Phi^{s}_j(x;z_0)  \,, \\
C^{(ij)}_{AP}(x_0, z_0) &=  \frac{a^6}{V} \sum_{\vec{x},  \vec{z} } \Tr \left[ S_i(x,z)^{\dag} \gamma_0 S_j(x,z)  \right] 
=  \frac{a^3}{N_s V} \sum_{\vec{x}}  \Phi^s_i(x;z_0)^{\dag} \gamma_0 \Phi^{s}_j(x;z_0)  \,, \\
C^{(ij)}_{PVP}(x_0,y_0;z_0) &= \frac{a^9}{V} \sum_{\vec{x}, \vec{y}, \vec{z}} 
\Tr \left[ S_i(x,z)^{\dag}  S_j(x,y)\gamma_0 S_j(y,z) \right] 
 \\ & \nonumber
=  \frac{a^3}{N_s V} \sum_{\vec{x}}  \widetilde{\Phi}_{ji}^{s \dag}(y;x_0,z_0) \gamma_0 \Phi^s_j(x;z_0)  \,,
\end{align}
\end{subequations}
with $V=L^3$ the spatial volume. We have used the $\gamma_5$-Hermiticity of the fermion propagator $S(x,y) = \gamma_5 S(y,x)^{\dag} \gamma_5$ and $\widetilde{\Phi}^s_{ji}$ is a sequential propagator given by $\widetilde{\Phi}^s_{ji}(y;x_0,z_0) = a^6\sum_{\vec x,\vec z} \gamma_5 S_j(y,x) \gamma_5 S_i(x,z) \eta_s(z)$. In practice, since the stochastic sources do not introduce a bias, the number of sources $N_s$ on each gauge configuration can be small. We choose $N_s=12$ such that the numerical cost would be the same if we used the usual point source method with a single source location. \\

To compute the correlation functions in Eqs.~(\ref{eq:dS}) and (\ref{eq:cor_cV}) we instead use point sources and the method of sequential propagators for the three-point correlation functions. A point source is first created on time slice $z_0$. Then, a sequential inversion is performed using the variation of the action between time slices $t_1$ and $t_2$ as a sequential source. We thereby have access to all $y_0$ values in the range $[t_1,t_2]$. To increase statistics, we also average over equivalent polarizations $k=1,2,3$. The values of $t_1$, $t_2$, $z_0$ and $y_0$ used in our simulations are summarized in Table~\ref{tab:cV_setup}. 
We have computed the correlation functions entering Eq.\ (\ref{eq:cor_cV}) to leading order in lattice perturbation theory (see Appendix~\ref{app:B}) in order to test our lattice QCD code.

\section{Results\label{sec:results}}

\subsection{Results for $\zv$, $\bv$ and $\bbarv^{\,\rm eff}$} 

\begin{figure}[t!]

	\includegraphics*[width=0.49\linewidth]{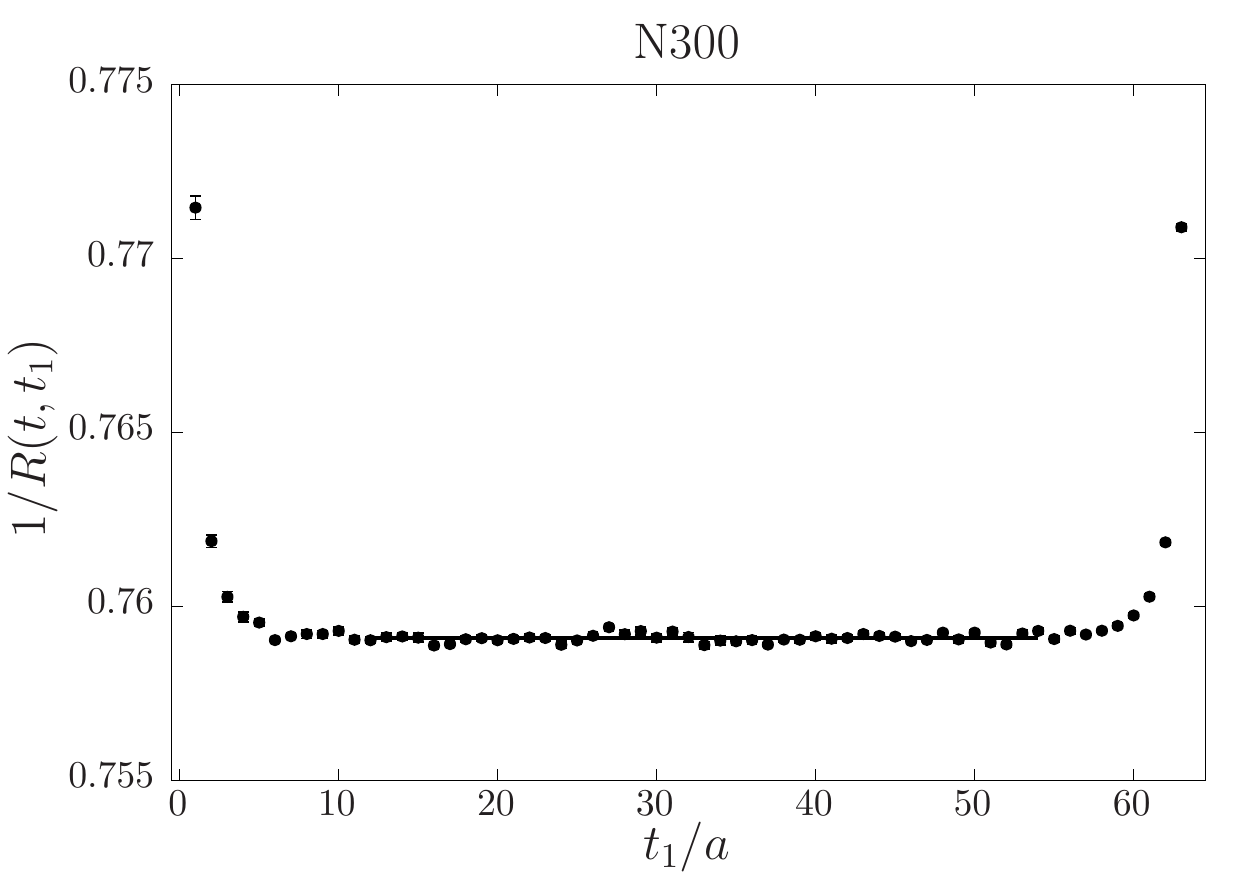}
	\includegraphics*[width=0.48\linewidth]{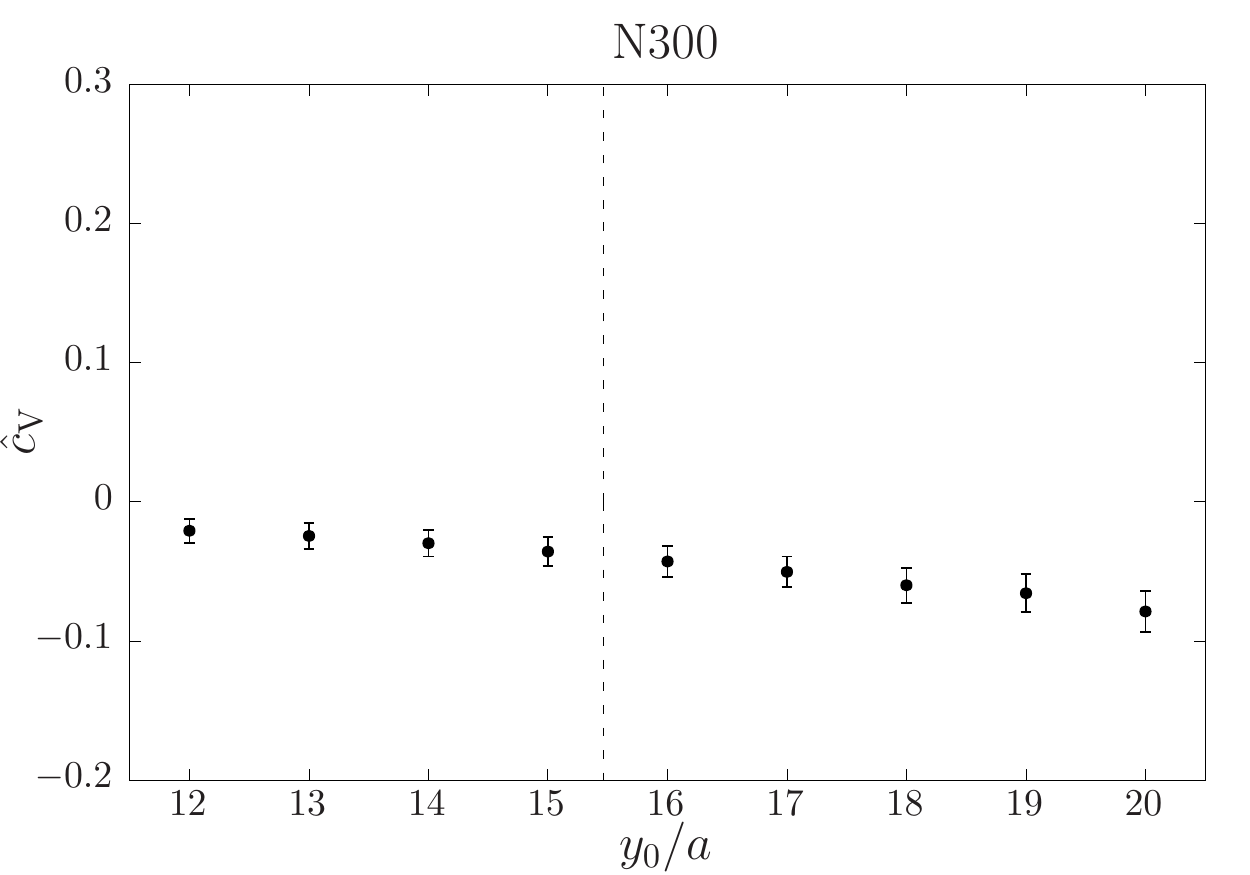}
		
	\caption{Plateaus for the ratio $R(t=\frac{T}{2},t_1)$ and $\hat{c}_{\rm V}$ defined through Eqs.~(\ref{eq:defZhat}) and (\ref{eq:cV_lat}) for the lattice ensemble N300.}
	\label{fig:ZV_plateaus}
\end{figure}

\begin{figure}[t!]
	\includegraphics*[width=0.49\linewidth]{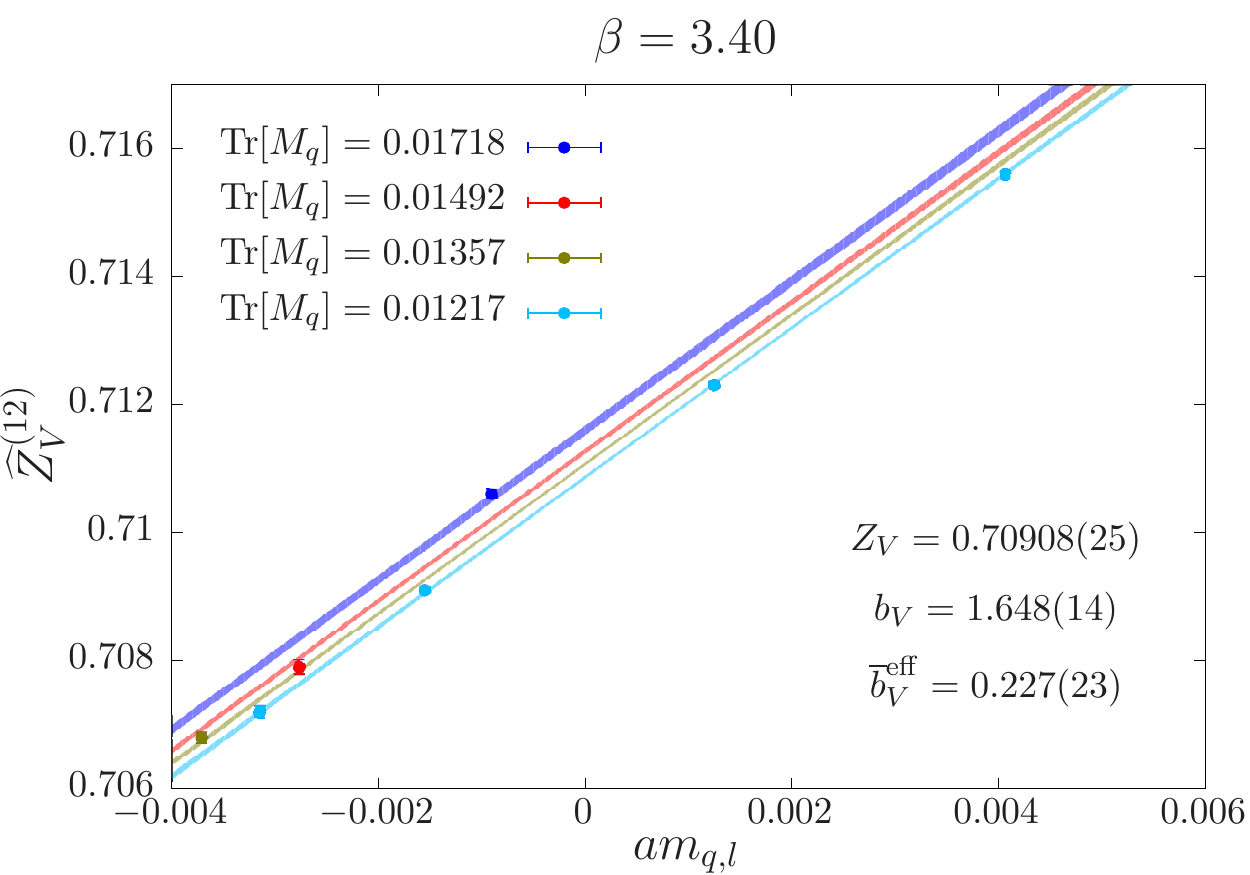}
	\includegraphics*[width=0.49\linewidth]{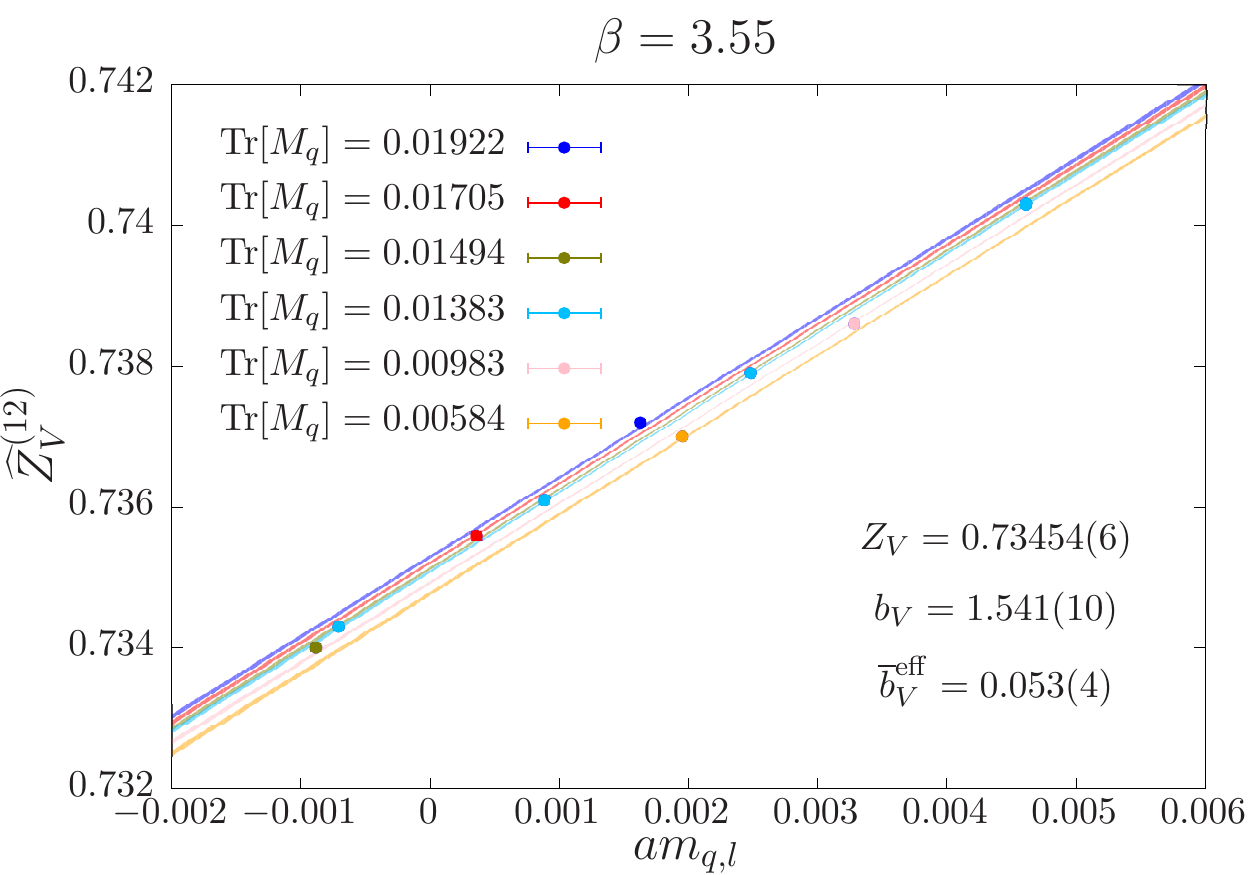}
	
	\caption{Results of the fits used to determine the renormalization constant $\zv$ and improvement coefficients $\bv$ and $\bbarv^{\,\rm eff}$ using the fit ansatz~(\ref{eq:ZV}) for two different values of the bare coupling.}	
	\label{fig:ZV_bV}
\end{figure}

Away from the boundary, it is convenient to use the variables $t=x_0-z_0$ and $t_1=y_0-z_0$. For each ensemble, the value of
$\hat{Z}_V$ is estimated from the ratio of three- to two-point correlation functions, defined through Eq.~(\ref{eq:condition}) with a local vector current. We choose $j=l$ (spectator quark), which define our renormalization scheme. The ratio has the asymptotic behavior
\begin{equation}
R(t, t_1) \xrightarrow[t_1,\;t-t_1\to\infty]{}   \frac{1}{\hat{Z}_V^{(12)}} \,,
\label{eq:defZhat}
\end{equation}
and is fitted to a constant in the plateau region where discretization effects are small. 
For ensembles with antiperiodic boundary conditions in time, we use $C_{PVP}(x_0,y_0;z_0) \to C_{PVP}(x_0,y_0;z_0) - C_{PVP}(x_0,y_0+T;z_0)$ to impose the vector Ward identity on each gauge configuration which can have a nonzero charge due to thermal fluctuations. 
Typical plots for the ensembles N200 and N300 are given in Fig.~\ref{fig:ZV_plateaus}, and the results for all ensembles are summarized in Table~\ref{tab:sim}. In a second step, the renormalization constant $\zv$, and the improvement coefficients $\bv$ and $\bbarv^{\,\rm eff}$ at a given value of the bare coupling $g_0$ are obtained using the fit ansatz 
\begin{equation}
\hat{Z}_{\rm V}^{(12)}(g_0^2,\mqav,m_{\rm q,12}) = \zv(g_0^2) \left(1 + 3\bbarv^{\,\rm eff}(g_0^2) a \mqav + \bv(g_0^2) am_{\rm q,12} \right) \,.
\label{eq:ZV}
\end{equation}
The ensembles included in the fit satisfy $|am_{q,l}| < 0.01$ and $a\mqav < 0.01$ such that higher order corrections are expected to be small. The results for each value of $\beta$ are given in Table~\ref{tab:res} and the statistical error includes the error on $\kappa_{\rm cr}$. The fits for two values of $\beta$ are shown in Fig.~\ref{fig:ZV_bV}. We note that the coefficient $\bv$ is significantly larger than the one-loop perturbative estimate given in Eq.~(\ref{eq:bvpert}) and that $\bbarv^{\,\rm eff} \ll \bv$. This was expected since the perturbative value starts only at two loops in perturbation theory. We provide the covariance matrices for the different values of the coupling considered in this work
\begin{subequations}
\begin{align} 
\mathrm{cov}(\zv, \bv, \bbarv^{\,\rm eff}; \beta=3.40) &= \begin{pmatrix}
	+6.04 \times 10^{-8} 	&	-1.05  \times 10^{-6}	 &	-5.02  \times 10^{-6}	\\
	-1.05  \times 10^{-6}	&	+1.93 \times 10^{-4}  &	+1.30 \times 10^{-4} \\
	-5.02  \times 10^{-6}	&	+1.30 \times 10^{-4} 	&	+5.50 \times 10^{-4} \\
	\end{pmatrix} \,, 
\label{eq:cov1} \\[1mm]
\mathrm{cov}(\zv, \bv, \bbarv^{\,\rm eff}; \beta=3.46) &= \begin{pmatrix}
	+4.16 \times 10^{-9} &	+9.92  \times 10^{-8}	 &	-6.68  \times 10^{-8}	\\
	+9.92  \times 10^{-8}	&	+1.90 \times 10^{-4}  &	-6.64 \times 10^{-5} \\
	-6.68  \times 10^{-8}	&	-6.64 \times 10^{-5} 	&	+2.97 \times 10^{-5} \\
	\end{pmatrix} \,, 
\label{eq:cov2} \\[1mm]
\mathrm{cov}(\zv, \bv, \bbarv^{\,\rm eff}; \beta=3.55) &= \begin{pmatrix}
	+3.17 \times 10^{-9} 	&	-2.85  \times 10^{-7}	 &	-1.27  \times 10^{-7}	\\
	-2.85  \times 10^{-7}	&	+9.21 \times 10^{-5}  &	+1.37 \times 10^{-5} \\
	-1.27  \times 10^{-7}	&	+1.37 \times 10^{-5} 	&	+1.37 \times 10^{-5} \\
	\end{pmatrix} \,,
\label{eq:cov3} \\[1mm]
\mathrm{cov}(\zv, \bv, \bbarv^{\,\rm eff}; \beta=3.70) &= \begin{pmatrix}
	+3.60 \times 10^{-9} &	+2.10  \times 10^{-7}	 &	-1.48  \times 10^{-7}	\\
	+2.10  \times 10^{-7}	&	+1.38 \times 10^{-4}  &	-6.14 \times 10^{-5} \\
	-1.48  \times 10^{-7}	&	-6.14 \times 10^{-5} 	&	+3.26 \times 10^{-5} \\
	\end{pmatrix} \,.
\label{eq:cov4}
\end{align}
\end{subequations}
Finally, we perform a Padé fit to obtain the renormalization factor and the improvement coefficients as a function of the bare coupling $g_0^2$. Our final results read
\begin{subequations}
\label{eq:padeZV}
\begin{align}
\zv(g_0^2) &= 1 - 0.10057\, g_0^2 \times \frac{1 + p_1 \, g_0^2 + p_2 \, g_0^4 }{1 + p_3 \, g_0^2 } \,, \\
\bv(g_0^2) &=  1 + 0.11813 \, g_0^2 \times \frac{1 + p_1  \, g_0^2 }{1 + p_2 \, g_0^2 }\,, \\
\bbarv^{\,\rm eff}(g_0^2)  &=  \frac{ p_1 \, g_0^4 }{ 1 + p_2 \, g_0^2} \,, 
\end{align}
\end{subequations}
which automatically reproduce the one-loop calculations and where the parameters and covariance matrices are given by  
\begin{subequations}
\begin{align} 
p(\zv) &= \begin{pmatrix}
	-0.2542 	\\
	-0.0961	\\
	-0.4796	\\
	\end{pmatrix} \,, \qquad 
\mathrm{cov}(\zv) = \begin{pmatrix}
	+1.31619 	&	+4.92750	&	+6.15758	\\
	+4.92750	&	+66.8321	&	+75.3218	\\
	+6.15758	&	+75.3218	&	+85.2733	\\
	\end{pmatrix} \times 10^{-6}\,, 
\label{eq:cov5} \\[1mm]
p(\bv) &= \begin{pmatrix}
	-0.184 	\\
	-0.444	\\
	\end{pmatrix} \,, \qquad \ \ \
\mathrm{cov}(\bv) = \begin{pmatrix}
	+36.7139 		&	+12.6698		\\
	+12.6698		&	+4.41224  	\\
	\end{pmatrix} \times 10^{-4}  \,, 
\label{eq:cov6} \\[1mm]
p(\bbarv^{\,\rm eff}) &= \begin{pmatrix}
	+0.00112 	\\
	-0.5577	\\
	\end{pmatrix} \,,  \qquad 
\mathrm{cov}(\bbarv^{\,\rm eff}) = \begin{pmatrix}
	+1.061463 	&	+14.53004	\\
	+14.53004	&	+248.5266 	 \\
	\end{pmatrix} \times 10^{-8} \,.
\label{eq:cov7} \
\end{align}
\end{subequations}

To ensure that O($a$) ambiguities vanish smoothly toward the continuum limit, the renormalization of the vector current must be performed along a line of constant physics (LCP). 
Since the CLS ensembles have different volumes, we checked explicitly that the impact on the renormalization factor is extremely small. 
The observable $\hat{Z}^{(12)}_{\rm V}$ has been computed on two sets of ensembles (H105/N101 and H200/N202) generated using the same lattice parameters but with different volumes and the results quoted in Table~\ref{tab:sim} are in perfect agreement within statistical errors.
Second, the correlation functions in Eq.~(\ref{eq:condition}) are computed with a source located at $z_0 = T/4$ to suppress boundary effects. For the ensemble N101, we have performed three sets of simulations with different source locations, $z_0 = T/4$, $T/4 - 4a$, $T/4 - 8a$, and the results are $\hat{Z}^{(12)}_{\rm V} = 0.70910(6)$, $0.70911(5)$ and $0.70919(6)$ respectively. The last results, where the source is close to the boundary, is slightly higher and might be affected by boundary effects. 
Those tests make us confident that with the procedure in Eq.~(\ref{eq:defZhat}) we indeed extract the matrix element in infinite volume.

\begin{figure}[t!]

	\includegraphics*[width=0.325\linewidth]{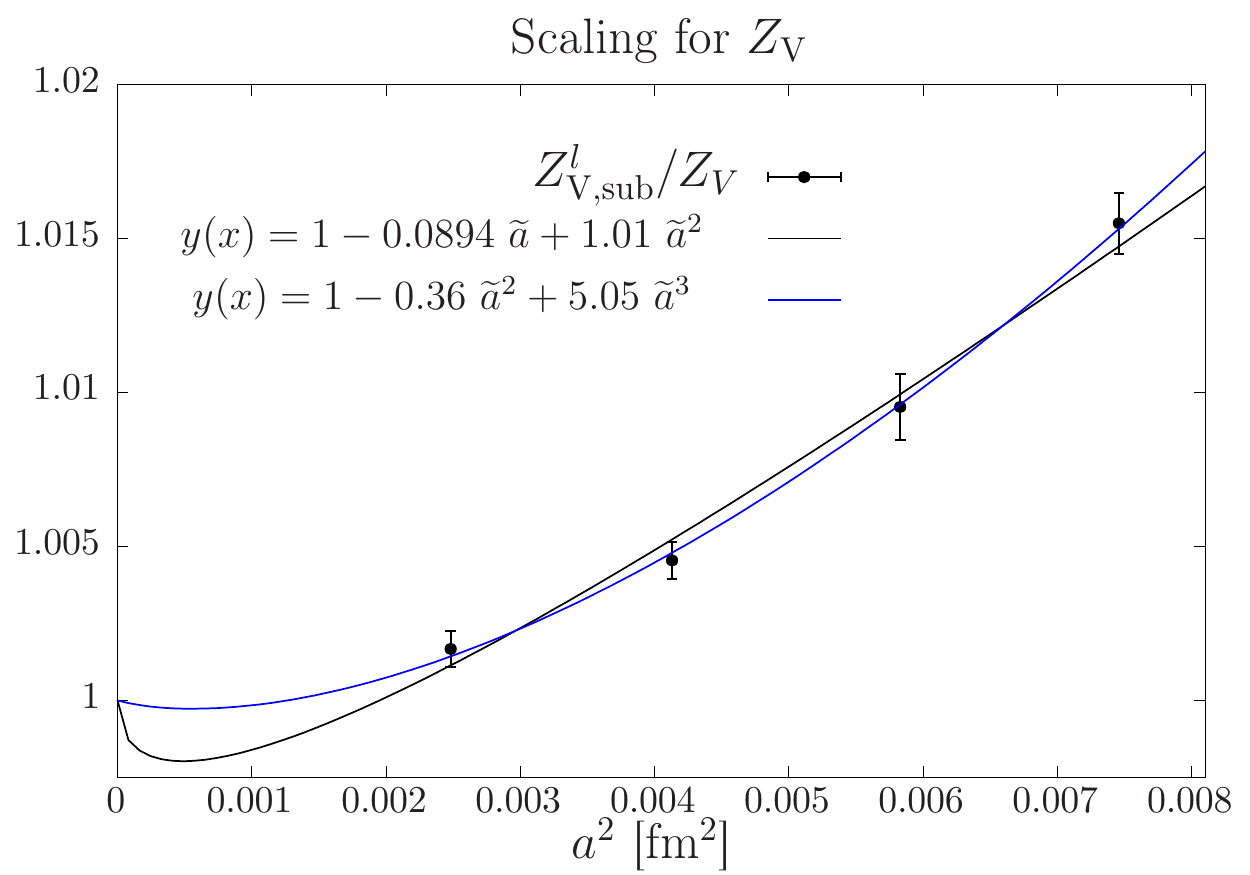}
	\includegraphics*[width=0.325\linewidth]{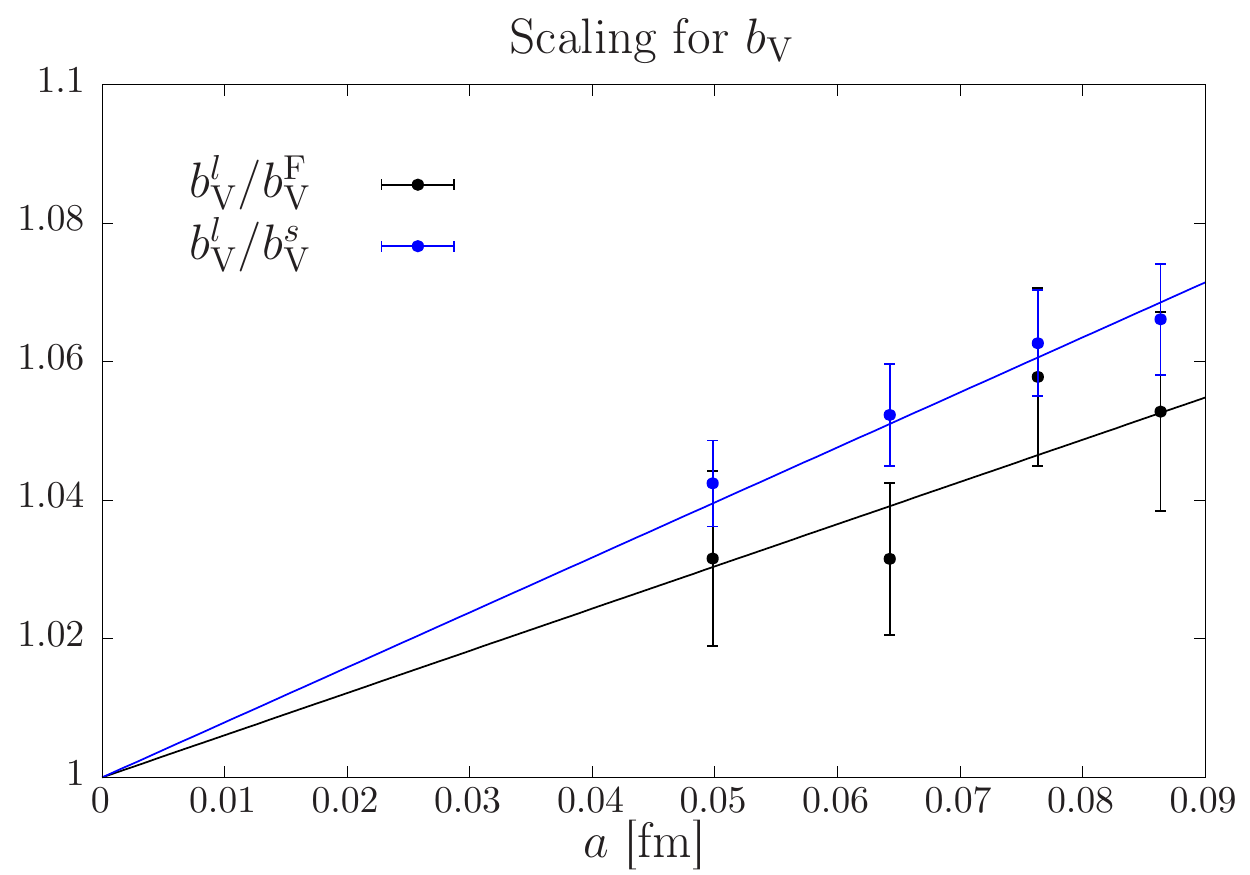}
	\includegraphics*[width=0.325\linewidth]{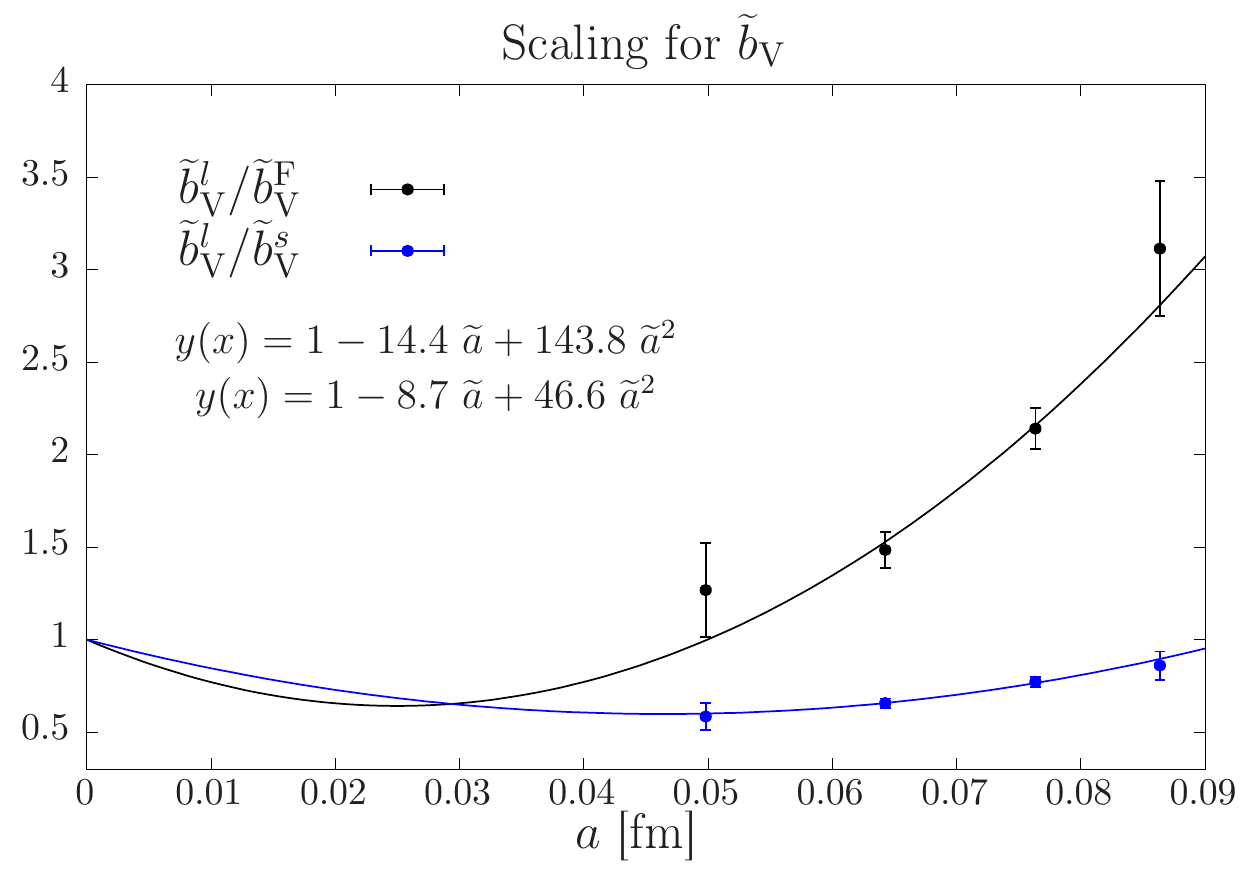}
	
	\caption{Left panel: Continuum limit of the ratio $Z_{\rm V,\, sub}^l/\zv$ where $Z_{\rm V,\, sub}^l$ was computed in~\cite{DallaBrida:2018tpn} and where $\zv$ refers to our own determination. Middle panel: $\bv^l$ and $\bv^s$ correspond to our determinations with the light or strange spectator quarks, respectively, and ${b}_{\rm V}^F$ is the value computed in~\cite{Fritzsch:2018zym}. Right panel: $\overline{b}_{\rm V}^F$ is the value computed in~\cite{Fritzsch:2018zym}. We have defined the dimensionless parameter $\widetilde{a} = a / 0.5~$fm. }
	\label{fig:scaling}
\end{figure}

As noted above, we could also choose $j=s$ for the spectator quark, and the values of the improvement coefficients $\bv$ and $\bbarv^{\,\rm eff}$ would differ by an O($a$)-ambiguity. To study this effect, we have repeated the analysis with $j=s$, and the results are given in Table~\ref{tab:res}. We do not observe any difference for the renormalization constant $\zv$ at our level of precision (both results should differ only by an O($a^2$) ambiguity). For $\bv$ and $\bbarv^{\,\rm eff}$, we observe variations by factors of at most $1.07$ and $1.7$, respectively. In Fig.~\ref{fig:scaling}, the continuum limit behavior of the ratio between the two results obtained using either a light or a strange spectator quark is shown in blue. For $\bv$, we observe the expected linear scaling with the lattice spacing. For $\bbarv$, the ratio goes to one only if one includes higher order discretization effects, which appear to be sizable.

\subsubsection{Comparison of results with previous work}

The renormalization factor $\zv$ has been determined independently in \cite{DallaBrida:2018tpn} using the chirally rotated Schr\"odinger functional framework. In Fig.~\ref{fig:scaling}, we plot the ratio $Z_{\rm V,\, sub}^l/\zv$ where $Z_{\rm V,\, sub}^l$ is extracted from \cite{DallaBrida:2018tpn} using the line of constant physics called $L_1$ and where the denominator corresponds to our own determination. This ratio goes rapidly to one in the continuum limit, even though the expected O$(a^2)$ scaling is not observed. However, the maximum deviation, obtained at $\beta=3.40$, is less than 1.6\%. Empirically, the available data for the departure of the ratio from unity can be described by the sum of a linear term and a quadratic term in the lattice spacing (not expected theoretically), or by the sum of a quadratic term and a cubic term. The latter fit in fact describes the data slightly better, see Fig.\ \ref{fig:scaling}. It also yields coefficients of reasonable size if one evaluates the lattice spacing say in units of 0.5\,fm.

\begin{figure}[t!]

	\includegraphics*[width=0.49\linewidth]{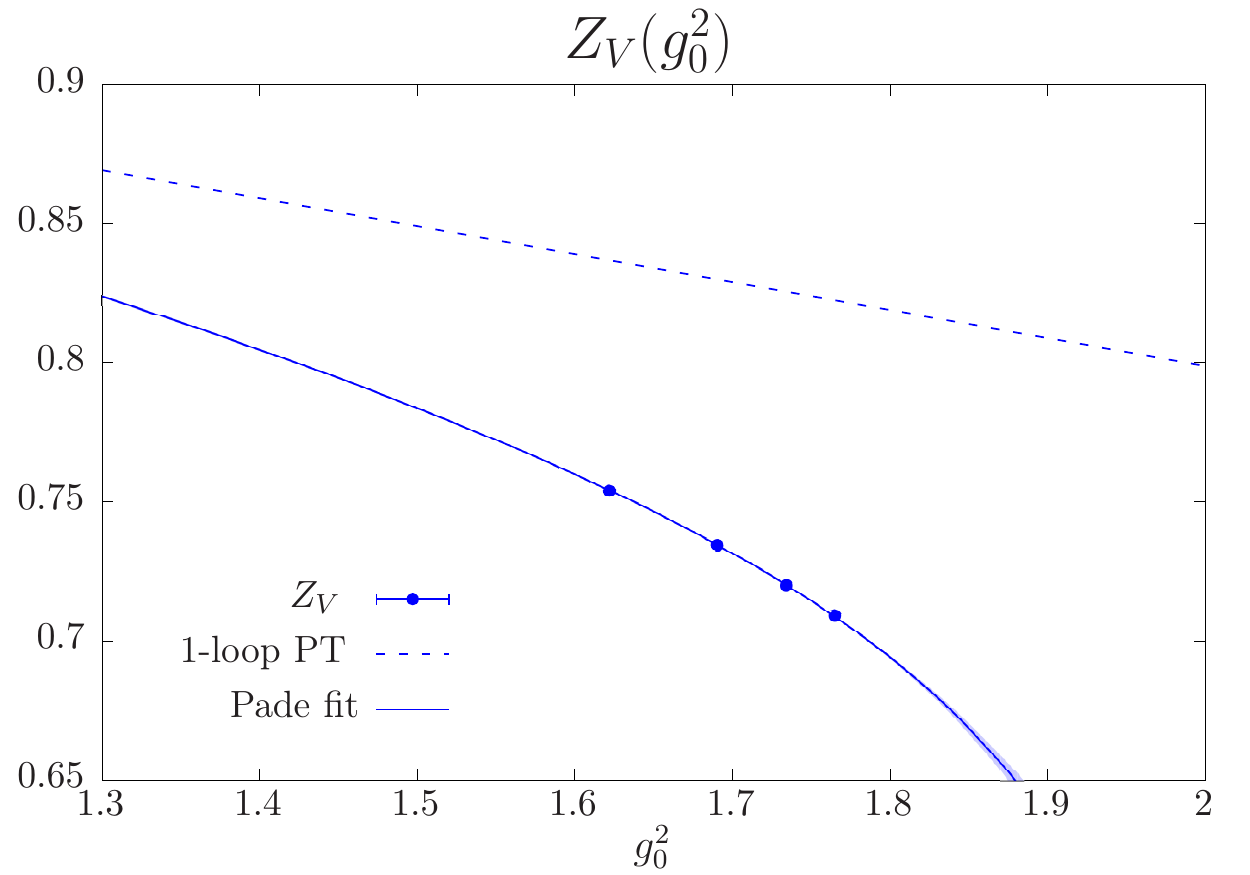}
	\includegraphics*[width=0.49\linewidth]{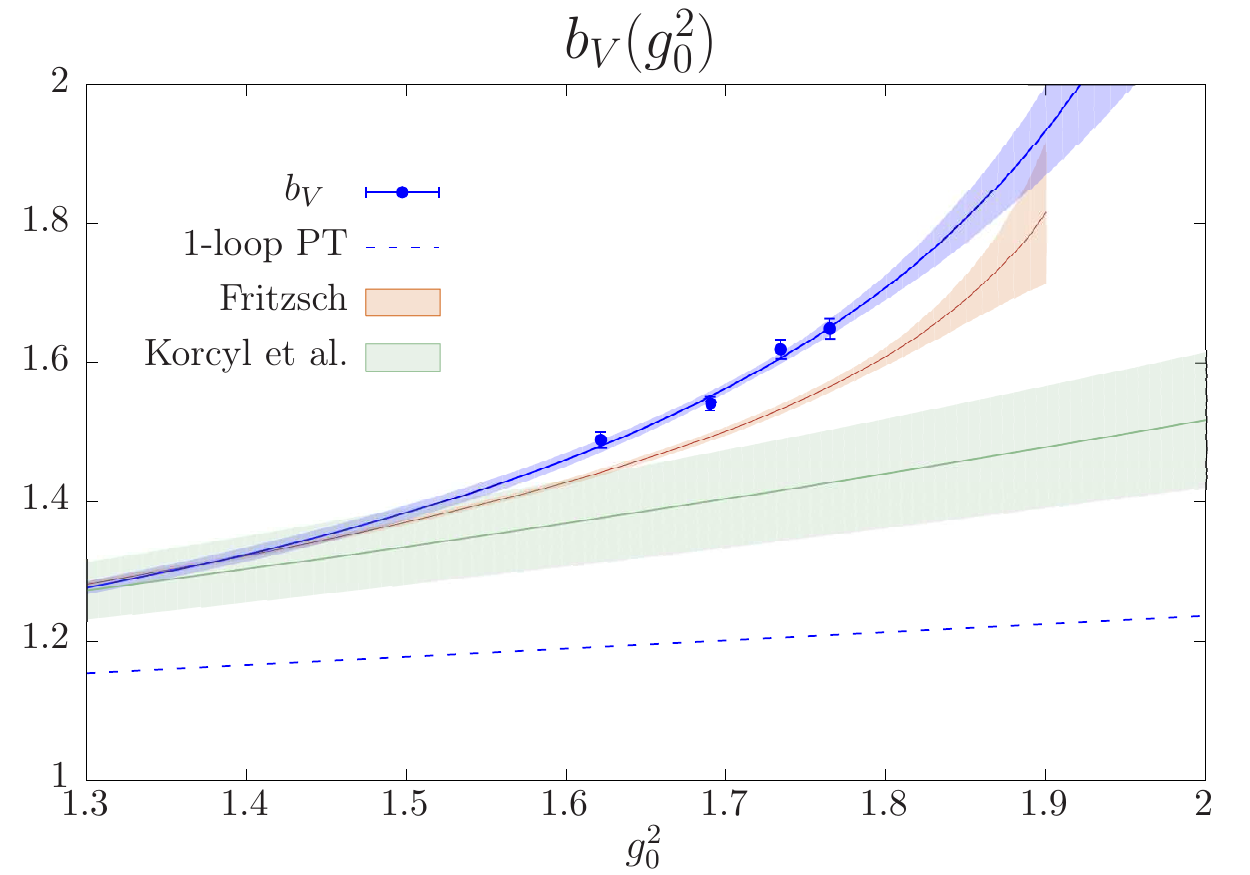}

	\includegraphics*[width=0.49\linewidth]{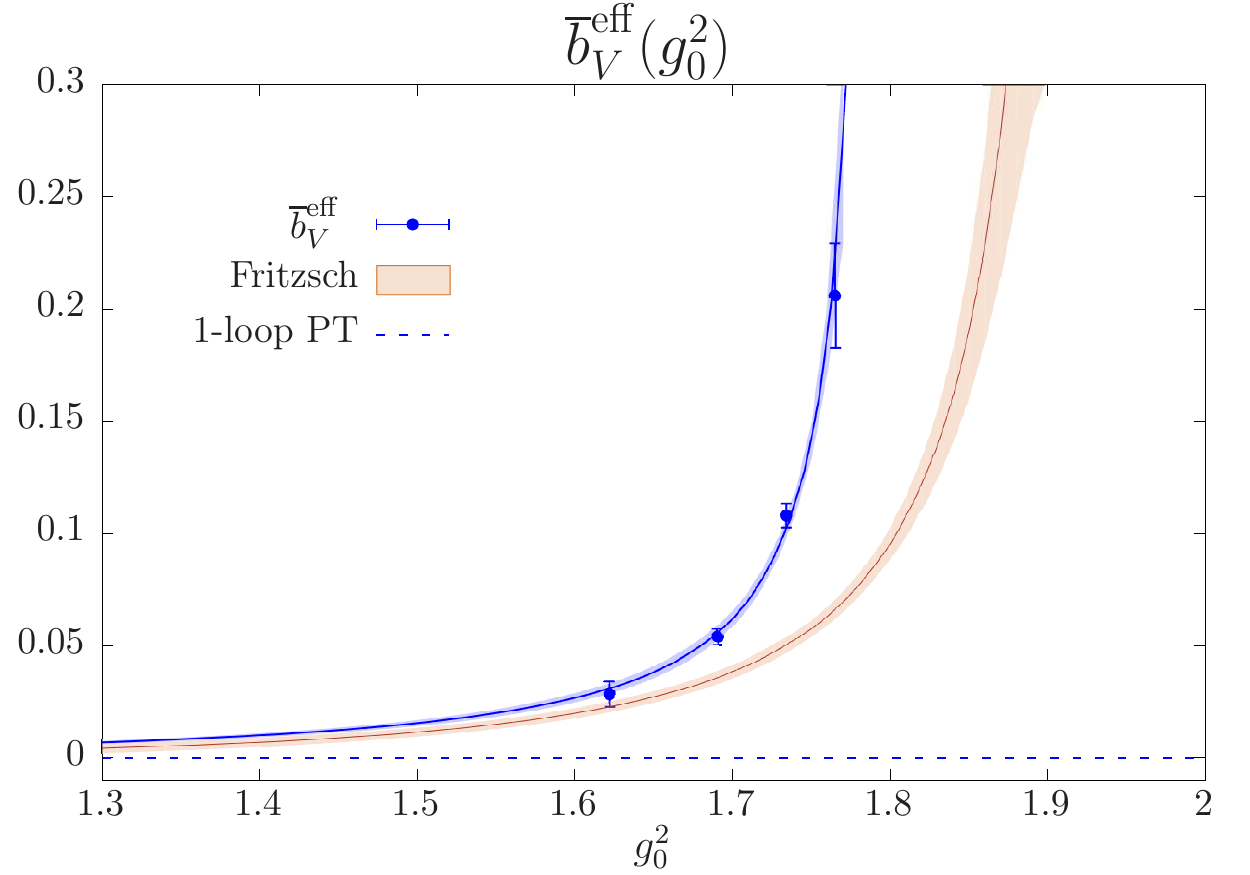}
	\includegraphics*[width=0.49\linewidth]{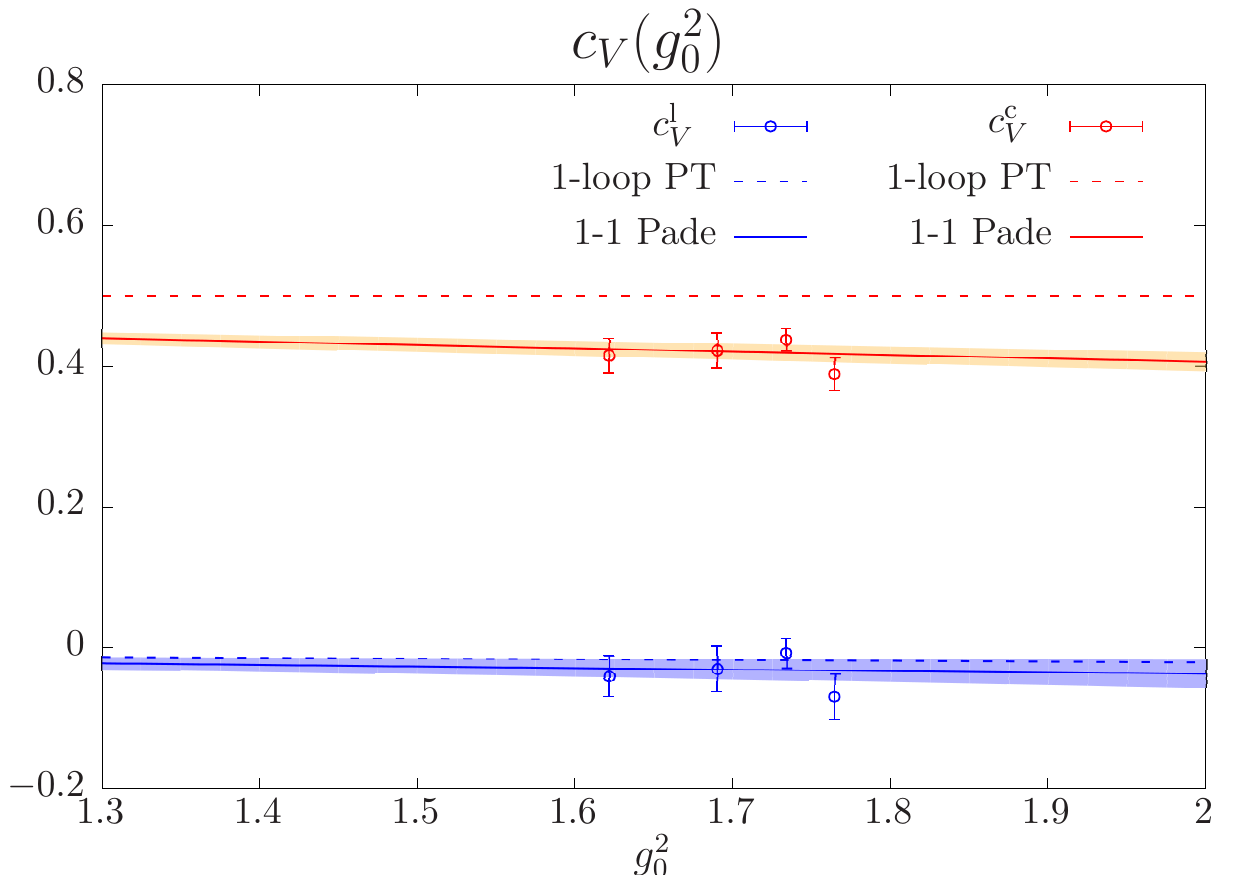}

	\caption{The dependence of the renormalization constant $\zv$
          and improvement coefficients $\bv$, $\bbarv^{\,\rm eff}$ and $\cv$ on
          the bare coupling $g_0^2$. The blue and red points
          correspond to the local and conserved vector currents
          respectively. The plain lines and error bands correspond to
          our Padé fits. For $\bv$ and $\bbarv^{\,\rm eff}$ we also compare our results with
          previous lattice
          determinations~\cite{Korcyl:2016ugy,Fritzsch:2018zym}. }
	\label{fig:ZV_bV_chiral}
\end{figure}

The coefficients $\bv$ and $\bbarv^{\,\rm eff}$ have also been determined recently in Ref.~\cite{Fritzsch:2018zym} using a different setup, based on the QCD Schr\"odinger functional. A comparison with our results is given in Fig~\ref{fig:ZV_bV_chiral}. For $\bv$, we observe a deviation of about 5\%, similar to the O($a$) dependence on the spectator-quark estimated above. In Fig.~\ref{fig:scaling}, we show the continuum limit behavior of the ratio with our own results and we observe the expected linear scaling.
However, for $\bbarv^{\,\rm eff}$, the difference with the results quoted in Ref.~\cite{Fritzsch:2018zym} is significant, especially at large couplings $g_0^2$. Again, as can be seen on Fig.~\ref{fig:scaling}, we do not observe a linear scaling in $a$ for the ratio of the two determinations, and higher order corrections cannot be neglected. It suggests that this parameter suffers from a large ambiguity. 

From a practical point of view, one should remember that a typical value of $a\mqav$ is 0.005 on the $\beta=3.55$ ensembles, so that with $3\bbarv^{\,\rm eff}\simeq 0.16$ even a $100\%$ ambiguity on $\bbarv$ has an impact below the permille level. 
Conversely, it could be that O($a^2$) effects compete with these terms, resulting in a substantial O($a$) contamination in our determination of $\bbarv^{\,\rm eff}$.
For the physics applications discussed in the Introduction, it is interesting to compare our values for the renormalization factor $\hat{Z}^{(12)}_{\rm V}$ of the isovector current to those one would obtain using the $\zv$ values of \cite{DallaBrida:2018tpn}
and the $\bv$ and $\bbarv^{\,\rm eff}$ values from \cite{Fritzsch:2018zym}. We find that our direct estimates are always slightly lower, and that the relative difference depends almost only on $g_0^2$: it is about 1.3\% at $\beta=3.40$, 0.8\% at $\beta=3.46$, 0.37\% at $\beta=3.55$ and 0.12\% at $\beta=3.70$. 
We conclude that these differences are of reasonable size, compatible with the expected $a^2$ (and higher order) ambiguity introduced by the choice of a specific renormalization condition.

\begin{figure}[t!]

	\includegraphics*[width=0.49\linewidth]{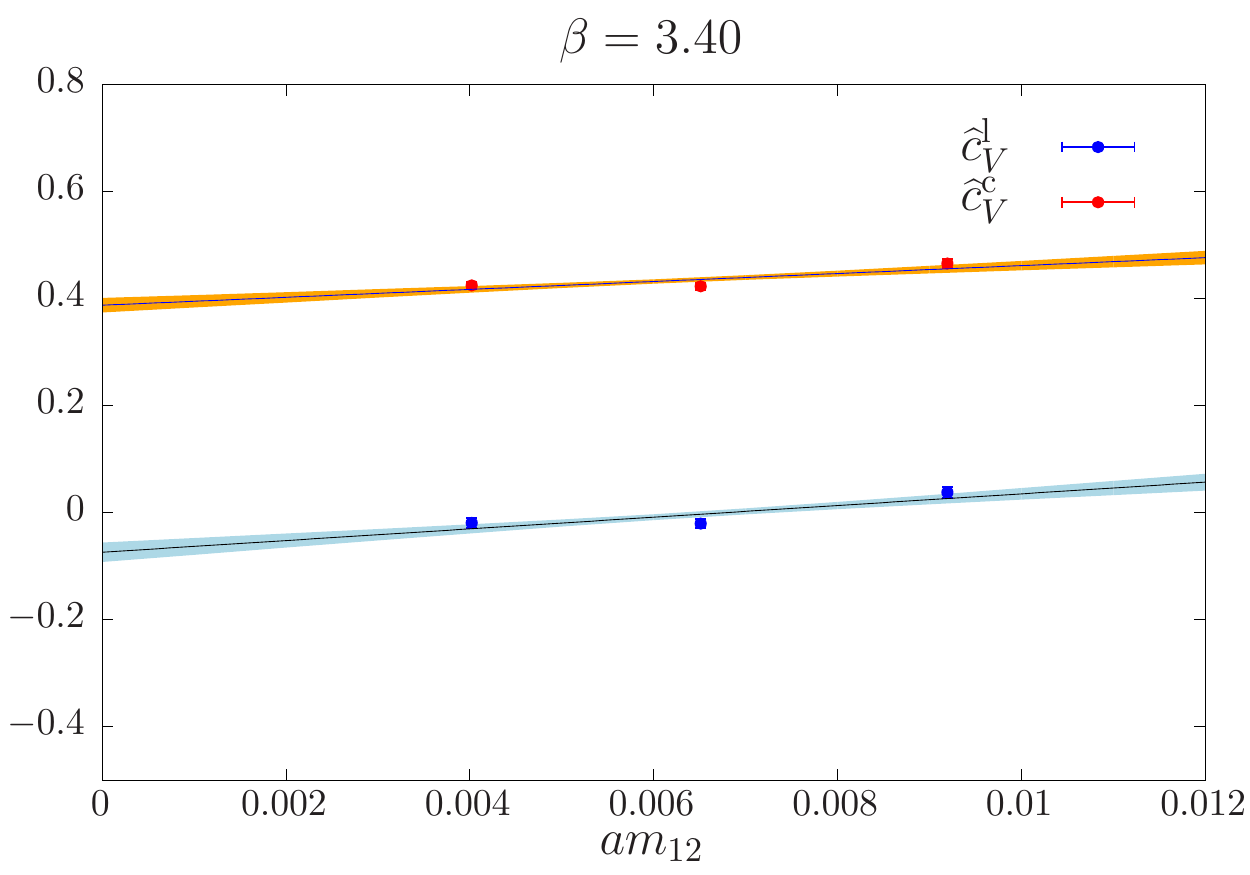}
	\includegraphics*[width=0.49\linewidth]{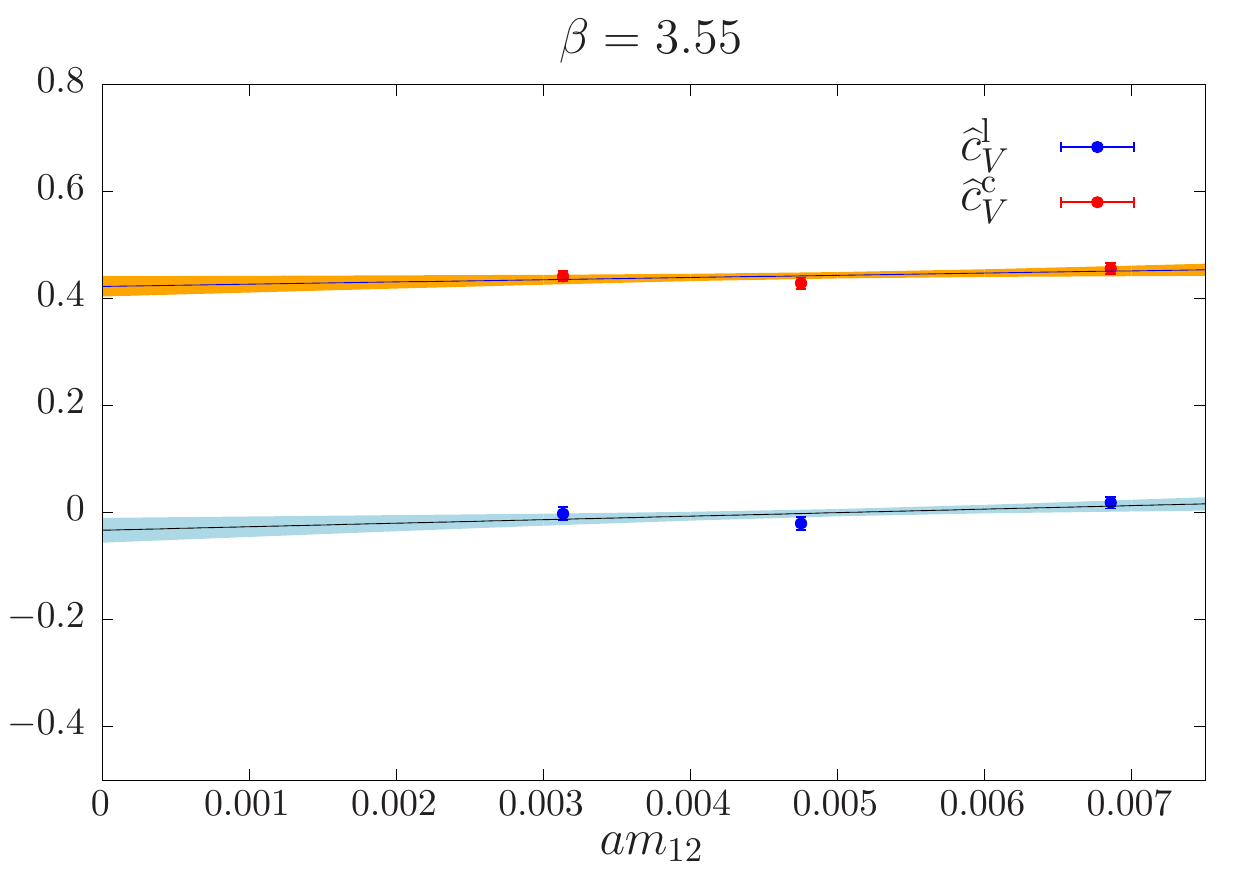}
		
	\caption{Chiral extrapolations of the improvement coefficients $\cv^{l}$ and $\cv^{c}$, respectively for the local and conserved vector currents, for two different values of the bare coupling.}	
	\label{fig:cV_extraps}
\end{figure}

\subsection{Results for the improvement coefficient $\cv$} 

For $y_0$ not too close from $t_1$ and $t_2$, we can extract the value of $\hat{c}_{\rm V}$ for each lattice ensemble. In practice, since we want to use a line of constant physics, we choose $y_0-z_0 = 0.77~\fm$ and interpolate linearly between two neighboring time slices when necessary. 
Deviations from LCP due to the different sizes of the lattices used should be very mild since we are working in large volumes.
The values of $z_0$, $t_1$ and $t_2$ in Eqs.~(\ref{eq:dS}) and (\ref{eq:cor_cV}) as well as the values of $y_0$ used in the interpolation are quoted in Table~\ref{tab:cV_setup}. 
Similar results are obtained using either the vector operator or the tensor operator as a probe operator in Eq.~(\ref{eq:cor_cV}). In practice, we use the linear combination $\mathcal{O}_{\rm ext}=V_{k}^{(31)}(z_0,\vec 0)+\Sigma_{k0}^{(31)}(z_0,\vec 0)$ which helps to improve the statistical precision. 
For $\za$, we use the results called $Z^l_{\rm A,sub}$ using the $L_1$-LCP from~\cite{DallaBrida:2018tpn}. For $\ba$ we used the published values in~\cite{Korcyl:2016ugy}, and for $\bbara$ we use values from~\cite{private} (in practice, we use $\overline{b}_{\rm A}^{\rm eff}$ which includes the $\bg$-term for $\za$, as in Eq.~(\ref{eq:bbareff_def})).
The results for $\hat{c}_{\rm V}$ for each ensemble are given in Table~\ref{tab:sim} and differ from $\cv$ by the presence of a contact term which vanishes in the chiral limit, as explained in Sec.~\ref{sec:cv}.  The improvement coefficient $\cv$ is obtained using a linear extrapolation in the light-quark PCAC mass $m_{12}$ at constant $\mqav$, and the results for each value of the bare coupling are summarized in Table~\ref{tab:res}. As can be seen in Fig.~\ref{fig:cV_extraps}, we observe a very mild chiral dependence.
The error is dominated by the statistical precision of the correlation functions and the error on $\bbara$.
The uncertainty on $\za$ and $\ba$ appears to have a negligible impact at our level of precision.
Finally, we perform linear or quadratic fits in $g_0^2$ to determine $\cv$ as a function of the bare coupling. The results for the local and the conserved vector currents read
\begin{subequations}
\label{eq:padecV}
\begin{align}
\cv^{l}(g_0^2) &=  -0.01030 \, C_{\rm F} \,g_0^2 \, \times \, (1 + 0.15(35) \, g_0^2) \,, \\ 
\cv^{cs}(g_0^2) &=  \frac{1}{2} \, \times \, (1 - 0.093(13) \, g_0^2) \,.
\end{align}
\end{subequations}
The parametrization is consistent with the perturbative predictions collected in Sec.~\ref{sec:pert}. Our values for $\cv^{l}$ are significantly smaller (in magnitude) than the preliminary values determined in \cite{Heitger:2017njs} by applying a similar improvement condition in the Schrödinger functional. It could be due to a large O($a$) ambiguity in the definition of $\cv$. However, our values for $\zv$ also differ by more that one standard deviation from the ones computed in \cite{Heitger:2017njs}. Since $\zv$ enters in the determination of $\cv$, this could partly explain the disagreement. For example, if we use the preliminary results for $\zv$ given in \cite{Heitger:2017njs}, our value would be $\hat{c}_{\rm V}^l =-0.146$ for H105 and $\hat{c}_{\rm V}^l = -0.178$ for N200. These observations highlight the need for good control over $\zv$ to determine precisely $\cv^l$. Since the point-split vector current is conserved, $\zv^c = 1$, this issue is absent for the determination of~$\cv^{cs}$.

\section{Conclusion\label{sec:concl}}

We have determined non-perturbatively the renormalization constant and improvement coefficients of the local and point-split nonsinglet vector currents with $\Nf = 2+1$ O($a$)-improved Wilson quark action and the tree-level Symanzik improved gauge action. Only one coefficient, $\fv$, is missing but is also expected to be small, as it starts at O$(g_0^6)$ in perturbation theory; in this regard, we note that for the two other mass-dependent improvement coefficients, $\bv$ and $\bbarv$, the hierarchy expected from perturbation theory is indeed observed in our non-perturbative results. All these parameters are required for the full O$(a)$-improvement of the vector current and the reduction of discretization effects in lattice simulations. They are essential in the calculation of the hadronic vacuum polarization (HVP) contribution to the muon $g-2$, where a precision below 1\% is aimed at in the near future. 

Full O$(a)$-improvement of the vector current requires one to consider the renormalization factor $\zv(\widetilde{g}_0)$ at the value of the renormalized coupling $\widetilde{g}_0$ instead of the bare coupling $g_0$. We have taken this difference into account by replacing the improvement coefficient $\bbarv$ by the effective parameter $\bbarv^{\,\rm eff}$, thus avoiding the use of the unknown coefficient $\bg$.

We have obtained the renormalization factor and improvement coefficients by imposing vector and axial Ward identities at finite lattice spacing and bare quark masses on a set of large volume ensembles.  
Deviations from the line of constant physics in our renormalization scheme have been studied and shown to be small for $\zv$, $\bv$ and $\bbarv$.

Our final results for the different $\beta$ values used in CLS simulations are summarized in Table~\ref{tab:res}. We also provide interpolating formulas through Eqs.~(\ref{eq:padeZV}) and (\ref{eq:padecV}).
As a cross-check of our methods, we have recomputed the improvement coefficient $\ca$ and find good agreement with the results of Ref.~\cite{Bulava:2015bxa}, which employ an improvement condition set up in the Schr\"odinger functional.

Our calculation is the first non-perturbative determination of the improvement coefficients $\cv^{\alpha}$ with $N_f=2+1$ Wilson quarks for both the local and (the symmetrized version of) the point-split vector currents. The value for the local vector current is small and both values, for the local and point-split vector currents, are close to their perturbative values. 

The comparison with the recent findings of Ref.~\cite{Fritzsch:2018zym} shows that a potentially large O($a$)-ambiguity in $\bbarv$ remains, but that it should vanish smoothly in the approach to the continuum limit.
For the vector current renormalization, we find important corrections to the expected asymptotic O($a^2$) scaling for the difference between our results and the recent determination of Ref.~\cite{DallaBrida:2018tpn}. However, we note the relative discrepancy is rather small, and smaller than that observed for two different normalization conditions for the axial current~\cite{DallaBrida:2018tpn,Bulava:2016ktf}.

In the future, other improvement coefficients may be determined for the lattice action used here, thanks to the availability of an extensive set of CLS lattice ensembles. In particular, the $\Nf=2+1$ hadronic contribution to the running of the weak mixing angle involves the flavor-singlet vector current, whose improvement coefficient $\bar c_{\rm V}$ is unknown. A method to determine the latter based on a chiral Ward identity was proposed in \cite{Bhattacharya:2005rb}.

\begin{table}[t]
\caption{Results for the renormalization constant $\zv$ and improvement coefficients $\bv$, $\bbarv^{\,\rm eff}$ and $\cv$ for different values of the bare coupling. For $\zv$, $\bv$ and $\bbarv^{\,\rm eff}$ the first (second) line corresponds to the results obtained with the light (strange) quark as a spectator quark.  For $\cv$, both results for the local and conserved vector currents are given. The value of critical hopping parameter $\kappa_{\rm cr}$ at $\beta = 3.40$ is extracted from \cite{Bali:2016umi}. }
\vskip 0.1in
 \begin{tabular}{l@{\hskip 02em}c@{\hskip 02em}c@{\hskip 02em}c@{\hskip 02em}c}
	\hline
	$\beta$			&		3.40		&	3.46			&	3.55			&	3.70	 \\
	\hline
	$\kappa_{\rm cr}$		& 	0.1369115(27)	&	0.1370645(10)  	&	0.1371726(13)	&	0.1371576(8)	\\ 
	\hline
	$\zv$			&	0.70908(25) 	&	0.71998(6)	&	0.73454(6)	&	0.75413(6)	\\
					&	0.70912(19)	&	0.71998(6)	&	0.73453(6)	&	0.75413(6)	\\
	$\bv$			&	1.648(14)		&	1.622(14)		&	1.541(10)		&	1.488(12) 	 	\\  
					&	1.546(10)		&	1.526(13)		&	1.460(09)		&	1.427(12)		\\
	$\bbarv^{\,\rm eff}$	&	0.206(23) 		&	0.108(05)		&	0.053(04)		&	0.029(06)		\\  
					&	0.240(17)		&	0.140(05)		&	0.081(04)		&	0.049(06)		\\
	$\bbarv$			&	0.227(23)		&	0.125(05)		&	0.067(04)		&	0.040(06) 		\\  
					&	0.260(17)		&	0.157(05)		&	0.095(04)		&	0.060(06) 		\\	
	\hline
	$\cv^{l}$			&	$-0.069(32)$	&	$-0.008(20)$	&	$-0.031(32)$	&	$-0.039(29)$	\\
	$\cv^{cs}$		&	0.389(23)		&	0.438(16)		&	0.422(26)		&	0.416(24)	\\
	\hline
 \end{tabular} 
\label{tab:res}
\end{table}

\begin{acknowledgments}
We thank Stefan Sint, Rainer Sommer and Mattia Dalla Brida for helpful discussions and Gunnar Bali and Piotr Korcyl for sharing with us preliminary results of the improvement coefficients of the axial current. We are grateful for the access to the ensembles used here, made available to us through CLS.  
Correlation functions were computed on the platforms ``Clover'' at the Helmholtz-Institut Mainz and ``Mogon~II'' at Johannes Gutenberg University Mainz. 
This work was partly supported by the European Research Council (ERC) under
the European Union's Horizon 2020 research and innovation programme through Grant Agreement no.\ 771971-SIMDAMA.
The CLS lattice ensembles used here were partly generated through computing time provided by the Gauss Centre for Supercomputing
(GCS) through the John von Neumann Institute for Computing (NIC) on the GCS share of the supercomputer JUQUEEN at J\"ulich Supercomputing Centre (JSC).
\end{acknowledgments}

\appendix

\section{Determination of the axial improvement coefficient $\ca$}
\label{app:C}

In this appendix, we use a similar setup to determine the improvement coefficient $\ca$ of the axial vector current (see
Eq.~(\ref{eq:imp_op})). The latter was previously determined non-perturbatively in~\cite{Bulava:2016ktf} in the framework of the
Schrödinger functional. This study can be seen as a consistency check of our method.

Within our numerical setup, described in Sec~\ref{sec:num_setup}, we can replace the axial vector current and the external operator $\mathcal{O}_{\ext}$ in Eq.~(\ref{eq:cor_cV}) by any other operator without new inversion of the Dirac operator. 
Therefore, we consider the following axial Ward identity
\begin{equation}
\int \mathrm{d}^3y \, \langle \delta S^{(12)} V_{R,0}^{(23)}(y_0,\vec{y}) \, \mathcal{O}_{\rm ext}^{(31)}(z_0,\vec 0) \rangle = \int \mathrm{d}^3y \, \langle A_{R,0}^{(13)}(y_0,\vec{y}) \, \mathcal{O}_{\rm ext}^{(31)}(z_0,\vec 0) \rangle \,,
\label{eq:cor_cA}
\end{equation}
with $\mathcal{O}_{\rm ext}^{(31)} = P^{(31)}$. The variation of the action is given by Eq.~({\ref{eq:dS}}), and, similarly to Eq.~(\ref{eq:cor_cV}), with the constraint $y_0 \notin [t_1,t_2]$. Here, $V_{R,0}^{(23)}$ and $A_{R,0}^{(13)}$ are the renormalized and O($a$)-improved vector and axial vector currents defined in Eq.~(\ref{eq:imp_op}).
If one knows $\zv$,  $\bv$, $\bbarv^{\,\rm eff}$,  $\ba$ and $\bbara$, then $\ca$ can be determined by imposing this equation to hold on the lattice up to O($a^2$) discretization effects, as done in Eq.~(\ref{eq:cV_lat}) for $\cv$.

\begin{figure}[t!]

	\includegraphics*[width=0.49\linewidth]{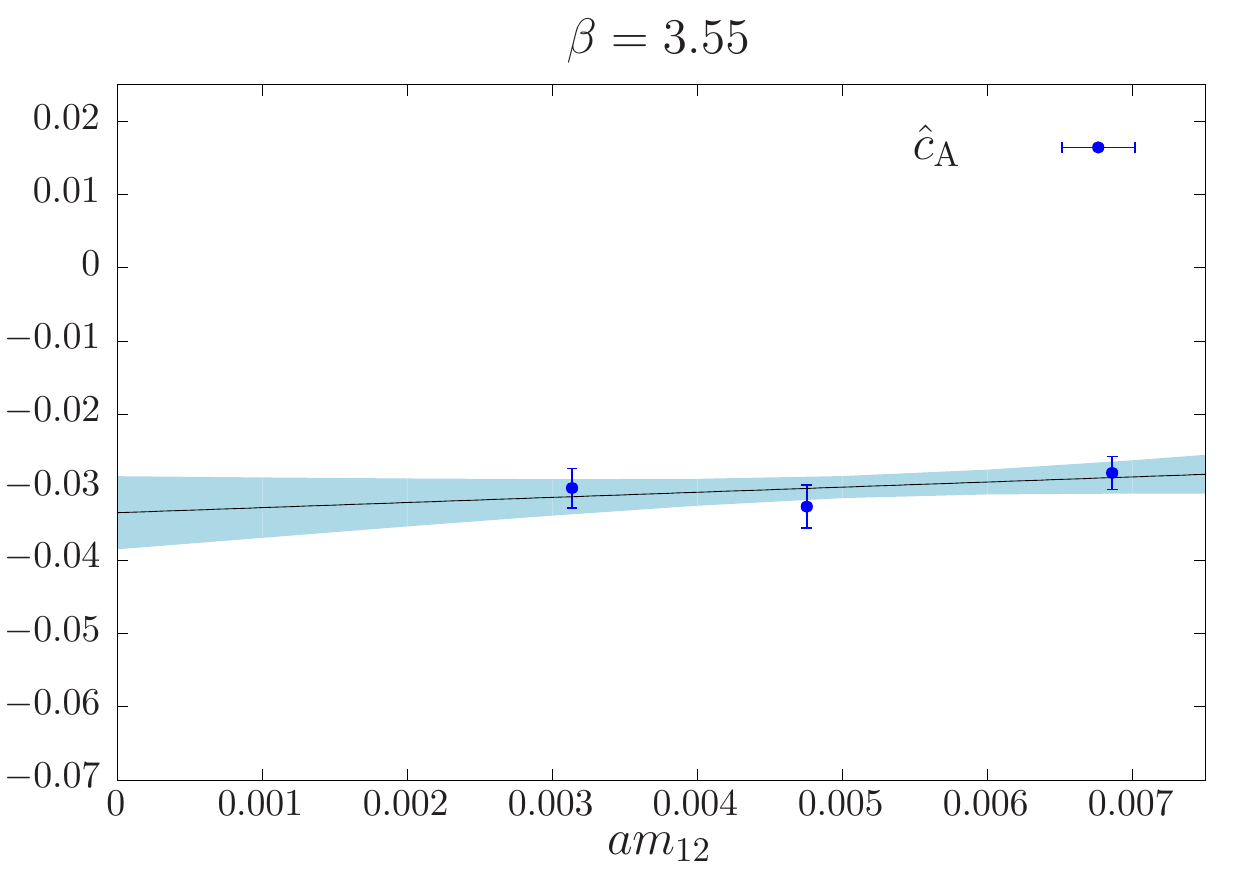}
	\includegraphics*[width=0.49\linewidth]{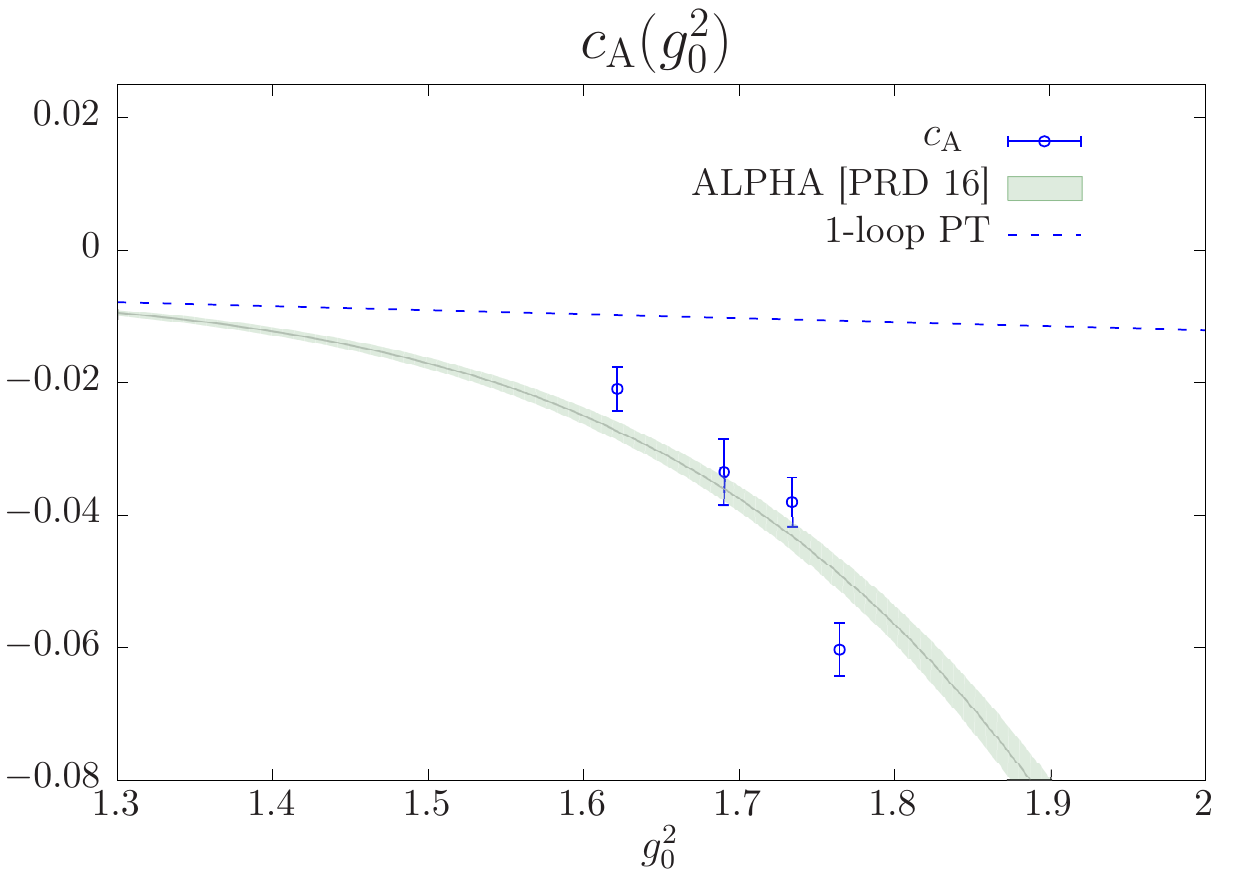}
		
	\caption{Left: Extrapolation of $\ca$ for one value of the bare coupling $g_0^2$. Right: Improvement coefficient $\ca$ as a function of the bare coupling $g_0^2$ with a (1,1)-Padé model. We also plot the results obtained by the ALPHA Collaboration in Ref.~\cite{Bulava:2016ktf} using a different method.}	
	\label{fig:cA_extraps}
\end{figure}

Using the same procedure as for $\cv$ and with the local vector current, we obtain the results summarized in Table~\ref{tab:sim} for a subset of the ensembles. Similar to $\cv$, the chiral dependence is very mild and the contribution from the contact term in the left-hand side of Eq.~(\ref{eq:cor_cA}) is removed by taking the limit $\mpcac_{12} \to 0$. We obtain the results quoted in Table~\ref{tab:res_fit_cA}. As can be seen on Fig.~\ref{fig:cA_extraps}, our results are close to the values quoted in~\cite{Bulava:2016ktf} obtained using a different
method. 
We point out that, in Eq.~(\ref{eq:cor_cA}), the variation of the action $\delta
S^{(12)}$ was computed with the value of $\ca$ published in
Ref.~\cite{Bulava:2016ktf} such that the two determinations are not
strictly independent.
Since this improvement coefficient has an O($a$)-ambiguity, we can attribute this small difference to the different schemes used.

\begin{table}[h!]
\caption{Results for the improvement coefficient $\ca$ at different values of the bare coupling.}
\vskip 0.1in
 \begin{tabular}{l@{\hskip 02em}c@{\hskip 02em}c@{\hskip 02em}c@{\hskip 02em}c}
	\hline
	$\beta$			&		3.40		&	3.46			&	3.55			&	3.70	 \\
	\hline
	$\ca$			&	$-0.060(4)$	&	$-0.038(4)$	&	$-0.033(5)$	&	$-0.021(3)$	\\
	\hline
 \end{tabular} 
\label{tab:res_fit_cA}
\end{table}

\section{Axial Ward identities in the free theory}
\label{app:B}

In the following, we give the tree-level expressions in lattice
perturbation theory for the correlation functions involved in the
chiral Ward identities, Eqs.~(\ref{eq:cor_cV}) and (\ref{eq:cor_cA}). 
We have used these expressions to test the lattice QCD code implementing the Ward identities.

We provide a more general expression in that we allow for a general spatial momentum,
on the other hand we restrict ourselves to the equal-mass case, $m_1=m_2=m_3=m$.
At order $g_0^0$ with a Wilson quark action, 
the correlation functions do not depend on $c_{\rm sw}$. 
We use the standard notation
\begin{eqnarray}
\hatp_\mu = \frac{2}{a} \sin\frac{ap_\mu}{2},\qquad \qquad 
\circp_\mu = \frac{1}{a} \sin ap_\mu,
\end{eqnarray}
as well as
\bea
A(\vec p) &=& 1 + am + \frac{1}{2} a^2 \hat{\vec p}^2,
\\
B(\vec p) &=& m^2 + (1+am) \hat{\vec p}^2 + \frac{1}{2} a^2 \sum_{k<l} \hatp_k^2 \hatp_l^2,
\\
C(\vec p) &=& m + \frac{a}{2} \Big({\hat{\vec p}^2} -\frac{B(\vec p)}{A(\vec p)}\Big) ,
\\
{\cal D}_{\vec p} &=&  \sqrt{B(\vec p)\,(4A(\vec p) + a^2 B(\vec p))} = \frac{2}{a}A(\vec p) \sinh(a\omega_{\vec p}).
\nonumber
\eea
Let $\pm i\omega_{\vec p}$ be the pole in $p_0$ of the fermion propagator. We note the identities
\bea
\frac{4}{a^2}\sinh^2(a\omega_{\vec p}/2) &=&  \frac{B(\vec p)}{A(\vec p)},
\\
\frac{1}{a^2}\sinh^2(a\omega_{\vec p}) &=& \vec{\circp}{}^2 + C(\vec p)^2.
\eea
The free fermion propagator in the time-momentum representation reads, with the convention ${\rm sign}(0)=0$,
\bea
S(x,y) &\equiv  & \<\psi(x) \bar\psi(y)\> = 
\int_{B} \frac{d^3\vec p}{(2\pi)^3} \frac{e^{-\omega_{\vec p}|x_0-y_0|+i\vec p\cdot(\vec x-\vec y)}}{{\cal D}_{\vec p}} 
\\ && \left({\rm sgn}(x_0-y_0) \frac{1}{a}\sinh(a\omega_{\vec p})\gamma_0 -i\vec\gamma\cdot\vec{\circp} 
 + C(\vec p) + \delta_{x_0,y_0}   \frac{1}{a} \sinh(a\omega_{\vec p})\right),
\nonumber
\eea
where $\int_B$ denotes integration over the Brillouin zone, $-\frac{\pi}{a}< p_i<\frac{\pi}{a}$.

In all three-point functions in this appendix, we assume $z_0<{\rm min}(y_0,x_0)$.
The correlation functions relevant to the Ward identity (\ref{eq:cor_cA}) are
\bea
&& a^3\sum_{\vec y}\; e^{-i\vec p\cdot (\vec y-\vec z)}\, \Big\<   A_0^{13}(y)\, P^{31}(z)\Big\> 
\\ && = 4 \,{\rm sign}(y_0-z_0) \int_B \frac{d^3q}{(2\pi)^3} \,
\frac{C(\vec q) \sinh\omega_{\vec p+\vec q} + C(\vec p+\vec q) \sinh(\omega_{\vec q})}{{\cal D}_{\vec q}{\cal D}_{\vec p+\vec q}}
e^{-|y_0-z_0|(\omega_{\vec p+\vec q}+\omega_{\vec q})},
\nonumber
\\
&& 
a^6 \sum_{\vec x,\vec z}\; e^{i\vec p\cdot(\vec z-\vec y)} \Big\<A_0^{12}(x)\,V_0^{23}(y)\,P^{31}(z) \Big\>
\label{eq:F0}
 \\ && \nonumber
=  8 \int_B\frac{d^3q}{(2\pi)^3}\, 
\frac{e^{-2\omega_{\vec q}\theta(x_0-y_0)(x_0-y_0) +(z_0-y_0)(\omega_{\vec p+\vec q}+\omega_{\vec q})}}
{{\cal D}_{\vec q}^2 {\cal D}_{\vec p+\vec q}}  f(\vec p,\vec q),\qquad 
\eea
with 
\be
f(\vec p,\vec q) = \left\{\begin{array}{l@{~~~}l} 
C(\vec q) \circvecq\cdot \circveck - C(\vec p+\vec q)\circvecq{}^2\Big|_{\vec k=\vec p+\vec q}   & x_0<y_0, 
 \\ 
C(\vec q)\Big(\sinh(\omega_{\vec q}) \sinh(\omega_{\vec p+\vec q}) + \circvecq\cdot \circveck 
     +C(\vec q)C(\vec p+\vec q) \Big)_{\vec k=\vec p+\vec q}   & x_0>y_0 \end{array}\right.
\ee
In Eq.~(\ref{eq:F0}), we have assumed $x_0\neq y_0$.
Finally, under the same assumptions we obtain
\bea
&& G_0(x_0-z_0,y_0-z_0,\vec p) 
\equiv a^6\sum_{\vec x,\vec z} e^{i\vec p\cdot(\vec z-\vec y)} \< P^{12}(x)\; V^{23}_0(y) P^{31}(z)\>
\\ && = -8\theta(x_0-y_0) \int_B\frac{d^3q}{(2\pi)^3}\, 
\frac{e^{-2\omega_{\vec q}\theta(x_0-y_0)(x_0-y_0)+(z_0-y_0)(\omega_{\vec p+\vec q}+\omega_{\vec q})}}
{{\cal D}_{\vec q}^2 {\cal D}_{\vec p+\vec q}}  \, g(\vec p,\vec q),
\nonumber
\eea
with 
\be
g(\vec p,\vec q)= \sinh(\omega_{\vec q})\Big(\sinh(\omega_{\vec q})\sinh(\omega_{\vec k}) + \circvecq\cdot \circveck
 + C(\vec q)C(\vec k) \Big)_{\vec k=\vec p+\vec q}.
\ee
The special case $x_0=y_0$ must be treated separately,
\bea
 G_0(y_0-z_0,y_0-z_0,\vec p) &=& -8 \int_B \frac{d^3q}{(2\pi)^3} \frac{e^{-(y_0-z_0)(\omega_{\vec p+\vec q}+\omega_{\vec q})}}
{{\cal D}_{\vec q}^2 {\cal D}_{\vec p+\vec q}}
\\ && \times \frac{\sinh(\omega_{\vec q})}{2}\Big([\sinh(\omega_{\vec q})+C(\vec q)][\sinh(\omega_{\vec k})+C(\vec k)]
 + \circvecq\cdot \circveck\Big)_{\vec k=\vec p+\vec q}.
\nonumber
\eea

The correlation functions relevant to the Ward identity Eq.\ (\ref{eq:cor_cV}) are 
\bea
&& a^3 \sum_{\vec y} e^{i\vec p\cdot(\vec z-\vec y)} \Big\< V_3^{l,13}(y) \; \Sigma_{30}^{31}(z) \Big\>
= -4 \;{\rm sign}(y_0-z_0)
\\ && \qquad \times\int_B \frac{d^3\vec q}{(2\pi)^3}\, 
\frac{e^{-(\omega_{\vec p} + \omega_{\vec p+\vec q})|z_0-y_0|}}{{\cal D}_{\vec q}{\cal D}_{\vec p+\vec q}} \, 
 \Big( C(\vec q)\sinh\omega_{\vec p+\vec q} + C(\vec p+\vec q) \sinh\omega_{\vec q} \Big) ,
\nonumber
\eea
and, for $x_0\neq y_0$ and $\vec k_\perp=(k_1,k_2)$, $\vec\ell_\perp=(\ell_1,\ell_2)$, 
\begin{eqnarray}
&& 
a^6 \sum_{\vec x,\vec z} e^{i\vec p\cdot (\vec z-\vec y)} \Big\<A_0^{12}(x) \; A_3^{23}(y)\;\Sigma_{30}^{31}(z)\Big\>
\\ &&= -16 \int_B \frac{d^3k}{(2\pi)^3}\, 
\frac{e^{-2\omega_{\vec k}(x_0-y_0)\theta(x_0-y_0)-(\omega_{\vec k}+\omega_{\vec p-\vec k})(y_0-z_0)}}
{{\cal D}_{\vec k}^2\, {\cal D}_{\vec p-\vec k}} \,f^A(\vec p,\vec k),
\nonumber \\
&& f^A(\vec p,\vec k) = 
\frac{1}{2}\Big\{\sinh(\omega_{\vec k})\Big( C(\vec k) \sinh(\omega_{\vec\ell})\theta(x_0-y_0)
   -\sinh(\omega_{\vec k}) C(\vec\ell) \theta(y_0-x_0)\Big)
\nonumber\\ &&
 -C(\vec k) \circk_3\circl_3  + C(\vec k) \circveck_\perp\cdot\circvecl_\perp 
+  C(\vec k)^2 C(\vec\ell) \Big\}_{\vec\ell=\vec p-\vec k}  ,
 \\ 
&& f^A(\vec 0,\vec k)= \left\{\begin{array}{l@{~~~}l} C(\vec k) (\circk_3{}^2 + C(\vec k)^2) & x_0>y_0  \\ 
-C(\vec k) \circveck_\perp{}^2  & x_0<y_0 \end{array}\right..
\end{eqnarray}
Further, for $x_0\neq y_0$, we have
\begin{eqnarray}
&&  G^A(x_0-z_0,y_0-z_0,\vec p) = 
a^6 \sum_{\vec x,\vec z}  e^{i\vec p\cdot(\vec z-\vec y)} \Big\<P^{12}(x) A_3^{23}(y)\Sigma_{30}^{31}(z)\Big\>
\\ && =  -16 \,\theta(x_0-y_0)\int_B \frac{d^3k}{(2\pi)^3}\, 
\frac{e^{-2\omega_{\vec k}(x_0-y_0)\theta(x_0-y_0)+(\omega_{\vec k}+\omega_{\vec p-\vec k})(z_0-y_0)}}
{{\cal D}_{\vec k}^2\, {\cal D}_{\vec p-\vec k}} \,g^A(\vec p,\vec k),
\nonumber
\\
&& g^A(\vec p,\vec k) = -\frac{1}{2}\sinh(\omega_{\vec k})
 \Big(\sinh(\omega_{\vec k})\sinh(\omega_{\vec\ell}) - \circk_3 \circl_3 + \circveck_\perp\cdot \circvecl_\perp 
 + C(\vec k)C(\vec\ell)\Big)_{\vec\ell=\vec p-\vec k},\qquad 
\\ && 
g^A(\vec 0,\vec k) = - \sinh(\omega_{\vec k}) (\circk_3{}^2 + C(\vec k)^2).
\end{eqnarray}
Finally, for $x_0=y_0$, 
\begin{eqnarray}
&&  G^A(y_0-z_0, y_0-z_0,\vec p) 
 =  -4 \int_B \frac{d^3k}{(2\pi)^3}\, 
\frac{e^{(\omega_{\vec k}+\omega_{\vec p-\vec k})(z_0-y_0)}}
{{\cal D}_{\vec k}^2\, {\cal D}_{\vec p-\vec k}} \,\sinh\omega_{\vec k}
\\ && \times\Big[ \circk_3 \circl_3 - \circveck_\perp\cdot \circvecl_\perp - C_{\vec\ell} C_{\vec k}
 - \sinh\omega_{\vec k}\sinh\omega_{\vec\ell} 
-C_{\vec\ell} \sinh\omega_{\vec k} - C_{\vec k} \sinh\omega_{\vec\ell}\Big]_{\vec\ell=\vec p-\vec k}.
\nonumber
\end{eqnarray}

\end{document}

%% file: table.tex
\begin{sidewaystable}

\caption{Parameters of the simulations: the bare coupling $\beta = 6/g_0^2$, the lattice resolution, the hopping parameter $\kappa$, the pion mass $m_{\pi}$ and the PCAC mass. $\hat{Z}_V^{(12)}$ is defined through Eq.~(\ref{eq:defZhat}): the two columns correspond to the light and strange spectator quarks respectively. The lattice spacing $a$ in physical units was determined in \cite{Bruno:2016plf}~: $a_{\beta=3.40} = 0.086~\fm$, $a_{\beta=3.46} = 0.076~\fm$, $a_{\beta=3.55} = 0.064~\fm$ and $a_{\beta=3.70} = 0.050~\fm$. }
\vskip 0.1in
\begin{tabular}{l@{\hskip 01em}lcl@{\hskip 01em}l@{\hskip 01em}l@{\hskip 01em}c@{\hskip 01em}l@{\hskip 01em}|@{\hskip 01em}l@{\hskip 01em}l@{\hskip 01em}l@{\hskip 01em}l@{\hskip 01em}l}
	\hline
Trajectory	&	CLS	&	$\quad\beta\quad$	&	$L^3\times T$ 	&	$\kappa_l$ &	$\kappa_s$	&	$m_{\pi}~[\MeV]$	&	$am_{\rm PCAC}$	&	$\hat{Z}_V^{(12)}$	&	$\hat{Z}_V^{(12)}$	&	$\hat{c}_V^l$ &	$\hat{c}_V^{cs}$	&	$\hat{c}_A$ \\
	\hline
\multirow{14}{*}{\shortstack{$m_q^{\rm av} =$~const. \\ with OBC}}	
	&	H101	&	$3.40$	&	$32^3\times96$	& 	0.13675962	&	0.13675962	& 	420 	&	0.009196(47)	&	0.71562(4) 	&	0.71562(4)	&	$+0.041(32)$		&	$+0.468(23)$	&	$-0.0404(25)$ \\  
	&	H102	&			&	$32^3\times96$	& 	0.136865		&	0.13654934	&	350	&	0.006512(49)	&	0.71226(6)	&	0.71252(5)	&	$-0.018(30)$		&	$+0.425(22)$	&	$-0.0505(20)$ \\  
	&	H105	&			&	$32^3\times96$	& 	0.136970		&	0.13634079	& 	280 	&	0.004021(55)	&	0.70908(5) 	&	0.70956(4)	&	$-0.017(30)$		&	$+0.426(22)$	&	$-0.0508(24)$\\  	
	&	N101	&			&	$48^3\times128$	& 	0.136970		&	0.13634079	& 	290	&	0.003994(32)	&	0.70911(5)	&	0.70955(5)	&	$\times$			&	$\times$		&	$\times$\\  	
	&	C101	&			&	$48^3\times96$	&	0.137030		&	0.13622204	&	220	&	0.002494(41)	&	0.70717(9)	&	0.70776(4)	&	$\times$			&	$\times$		&	$\times$\\
	\cline{2-13}
	&	S400		&	$3.46$	&	$32^3\times128$	&	0.136984		&	0.13670239	&	350	& 	0.005673(36)	&	0.72355(5)  	&	0.72372(5)	&	$+0.029(18)$		&	$+0.464(13)$	&	$-0.0354(26)$\\
	&	N401	&			&	$48^3\times128$	&	0.1370616	&	0.13654808	&	290	&	0.003781(26)	&	0.72116(5)	&	0.72148(5)	&	$-0.004(15)$		&	$+0.443(13)$	&	$-0.0406(21)$\\
	\cline{2-13}
	&	H200	&	$3.55$	&	$32^3\times96$	& 	0.137000		&	0.137000		& 	420	&	0.006858(23)	&	0.74028(5)	&	0.74028(5) 	&	$+0.021(30)$		&	$+0.458(22)$	&	$-0.0279(22)$\\  
	&	N202	&			&	$48^3\times128$	& 	0.137000		&	0.137000		& 	410  	&	0.006856(16)	&	0.74028(4)	&	0.74028(4)	&	$\times$			&	$\times$		&	$\times$\\
	&	N203 	&			&	$48^3\times128$	&	0.137080		&	0.13684028	& 	340 	&	0.004761(13)	&	0.73792(4)	&	0.73802(5)	&	$-0.018(29)$		&	$+0.431(23)$	&	$-0.0327(29)$\\ 
	&	N200 	&			&	$48^3\times128$	& 	0.137140		&	0.13672086	& 	280	&	0.003145(15)	&	0.73614(6) 	&	0.73632(4)	&	$-0.002(28)$		&	$+0.442(22)$	&	$-0.0303(25)$\\ 
	&	D200 	&			&	$64^3\times128$	& 	0.137200		&	0.13660175	& 	200 	& 	0.001538(12)	&	0.73429(4)	&	0.73461(4)  	&	$\times$			&	$\times$		&	$\times$	\\ 
	\cline{2-13}
	&	N300	&	$3.70$	&	$48^3\times128$	&	0.137000		&	0.137000 	   	&	420	&	0.005509(09)	&	0.75909(2)	&	0.75909(2)	&	$-0.047(25)$		&	$+0.409(20)$	&	$-0.0205(18)$ \\
	&	N302	&			&	$48^3\times128$	&	0.137064		&	0.13687218 	&	340	& 	0.003712(11)	&	0.75719(3)	&	0.75728(3) 	&	$-0.063(26)$		&	$+0.399(20)$  	&	$-0.0213(25)$\\
	&	J303		&			&	$64^3\times192$	&	0.137123		&	0.13675466 	&	260	& 	0.002047(08)	&	0.75544(2)	&	0.75557(3)	&	$-0.036(28)$		&	$+0.418(24)$	&	$-0.0206(33)$\\
	\hline
 	\hline
\multirow{4}{*}{\shortstack{$m_{q,s} = m_{q,s}^{\rm phys}$ \\ with OBC}}	
	&	H107	&	$3.40$	&	$32^3\times96$	&	0.13694567	&	0.13620317	& 	360	& 	0.006731(81) 	&	0.71056(8)  	&	0.71110(6)	&	$\times$			&	$\times$	&	$\times$\\  	
	&	H106	&			&	$32^3\times96$	& 	0.13701557	&	0.13614870	& 	280	& 	0.004111(73) 	&	0.70792(12) 	&	0.70853(6)	&	$\times$			&	$\times$	&	$\times$ \\  	
	&	C102	&			&	$48^3\times96$	&	0.13705085	&	0.13612906	&	230	&	0.002573(42)	&	0.70676(8)	&	0.70731(6)	&	$\times$			&	$\times$	&	$\times$\\
	\cline{2-13}
	&	N204	&	$3.55$	&	$48^3\times128$	& 	0.137112		&	0.13657505	& 	350 	&	0.004884(26)	&	0.73715(4)	&	0.73745(4)	&	$\times$			&	$\times$	&	$\times$	\\  
	&	N201 	&			&	$48^3\times128$	& 	0.13715968	&	0.13656132	& 	280	&	0.003166(34)	&	0.73564(6) 	&	0.73600(6)	&	$\times$			&	$\times$	&	$\times$	 \\ 
	&	D201 	&			&	$64^3\times128$	& 	0.1372067	&	0.13654684	&	200 	& 	0.001526(25)	&	0.73401(5)	&	0.73436(5)	&	$\times$			&	$\times$	&	$\times$	\\ 
	\cline{2-13}
	&	N304	&	$3.70$	&	$48^3\times128$	&	0.13707933	&	0.13666543	&	350	& 	0.003603(43)	&	0.75683(6)	&	0.75702(6)	&	$\times$			&	$\times$	&	$\times$	\\
	\hline
	\hline
\multirow{6}{*}{\shortstack{$m_{q,l} = m_{q,s}$\\ with PBC}} 
	&	rqcd29	&	$3.46$	&	$32^3\times64$	&	0.136600		&	0.136600		&	710 	&	0.021687(77)	&	0.73741(6) 	&	0.73741(6)	&	$\times$ 			& 	$\times$	&	$\times$\\
	&	B450		&			&	$32^3\times64$	&	0.136890		&	0.136890		&	410 	&	0.008071(36) 	&	0.72647(2)	&	0.72647(2)	&	$+0.020(16)$		&	$+0.455(12)$	&	$-0.0403(11)$\\
	&	rqcd30	&			&	$32^3\times64$	&	0.1369587	&	0.1369587	&	320 	&	0.004832(59)	&	0.72398(6) 	&	0.72398(6) 	&	$\times$			&	$\times$	&	$\times$\\
	&	X450		&			&	$48^3\times64$	&	0.136994		&	0.136994		&	260 	&	0.003305(31)	&	0.72265(3)	&	0.72265(3)	&	$\times$			&	$\times$	&	$\times$\\
	\cline{2-13}
	&	X250		&	$3.55$	&	$48^3\times64$	&	0.137050		&	0.137050		&	350 	&	0.004919(23) 	&	0.73863(2)	&	0.73863(2)	&	$\times$			&	$\times$	&	$\times$	\\
	&	X251		&			&	$48^3\times64$	&	0.137100		&	0.137100		&	270	&	0.002877(26)	&	0.73698(2)	&	0.73698(2)	&	$\times$			&	$\times$	&	$\times$	\\
	\cline{2-13}
	\hline
	\hline
\multirow{2}{*}{\shortstack{$m_{q,l} = m_{q,s}$\\ with OBC}}	
	&	H401	&	$3.46$	&	$32^3\times96$	&	0.136725		&	0.136725		&	590	&	0.015694(67)	&	0.73260(8)	&	0.73260(8)	&	$\times$			&	$\times$	&	$\times$	\\
	&	H400	&			&	$32^3\times96$	&	0.13688848	&	0.13688848	& 	420	&	0.008235(81)	&	0.72634(10)	&	0.72634(10)	&	$\times$			&	$\times$	&	$\times$	\\
	\cline{2-13}
	&	N303	&	$3.70$	&	$48^3\times128$	&	0.136800		&	0.136800		&	650 	&	0.012518(20)	&	0.76544(2)	&	0.76544(2)	&	$\times$			&	$\times$	&	$\times$	\\
	&	N301	&			&	$48^3\times128$	&	0.137005		&	0.137005		& 	420 	&	0.005334(19)	&	0.75897(3)	&	0.75897(3)	&	$\times$			&	$\times$	&	$\times$	\\

\hline
 \end{tabular} 
 \vskip 0.1in
\label{tab:sim}

\end{sidewaystable}